\documentclass[a4paper,11pt]{article}
\usepackage[left=2cm,right=2cm,top=2cm,bottom=3cm]{geometry}
\usepackage[utf8]{inputenc}

\usepackage{graphicx}
\usepackage{epstopdf,epsfig}

\AppendGraphicsExtensions{.tiff}
\PrependGraphicsExtensions{.tiff}

\usepackage{hyperref}
\usepackage{xcolor}
\usepackage{pdfpages}
\usepackage[british]{babel}
\usepackage{adjustbox}
\usepackage{physics}  
\usepackage{caption}
\usepackage{subcaption}
\usepackage{sidecap}
\usepackage[round]{natbib}
\usepackage{enumitem}
\usepackage{authblk}
\usepackage{lscape}
\usepackage{xfrac}
\usepackage[version=4]{mhchem}
\usepackage{amsmath,amssymb,stmaryrd}
\SetSymbolFont{stmry}{bold}{U}{stmry}{m}{n}

\graphicspath{{Figures/}}

\usepackage{printlen}
\usepackage{booktabs}            
\usepackage{array}

\newcommand{\HYPRE}{{\it hypre}}

\title{A 2D fluid model in BOUT++/Hermes-3 for simulating detached divertor plasmas in tokamaks}

\author[1]{Malamas Tsagkaridis}

\author[1]{Ben Dudson\footnote{Email: dudson2@llnl.gov}}
\author[2]{Mike Kryjak}
\author[1]{Nami Li}
\author[1]{Xueqiao Xu}

\affil[1]{Lawrence Livermore National Laboratory, 7000 East Avenue, Livermore, CA 94550, USA}
\affil[2]{United Kingdom Atomic Energy Authority, Culham Campus, Abingdon, OX14 3DB, UK}

\date{}
\begin{document}

\maketitle





\section*{Abstract} \label{sec:Abstract}

Detached divertor operation is essential for controlling target heat fluxes in future high-power tokamaks. We present a two-dimensional fluid model implemented in BOUT++/Hermes-3 for simulating detached divertor plasmas with plasma-neutral interactions and fixed-fraction impurity radiation. 
We report improvements to the PETSc SNES-based steady-state solver used for two-dimensional axisymmetric transport simulations with fluid neutrals and assess the performance of several sparse-solver preconditioning strategies. For the cases considered here, STRUMPACK used as an ILU preconditioner with low-rank compression provides the most favorable performance, reducing the linear-to-nonlinear iteration ratio and achieving approximately a fivefold speedup relative to \HYPRE{}-based configurations. The optimized solver is then applied to DIII-D shot $\#174287$ at 4000~ms to characterize detached divertor conditions. Cross-field transport coefficients are adjusted to reproduce upstream experimental profiles, and detachment is modeled using nitrogen fixed-fraction radiation. Steady-state particle and energy balances are evaluated by decomposing the density and pressure equations into their transport and source terms. In the outer divertor, electron energy losses are dominated by ionization and impurity radiation, while ion energy losses occur mainly through collisional transfer to electrons. An upstream separatrix density scan shows decreasing target electron temperature and a rollover of the integrated target ion saturation current at higher $n_{e,\mathrm{sep}}$ consistent with expected detachment trends. These developments establish a foundation for future multispecies simulations. Hermes-3 is publicly available on GitHub under the GPL-3 open-source license.

\bigskip
\noindent \textbf{Keywords:} tokamak, plasma edge modeling, detachment, Hermes-3.


\section{Introduction} \label{sec:Introduction}

Plasma detachment is a key operational regime for magnetic-fusion devices because it reduces the particle and heat fluxes incident on divertor targets. In future reactors, the target heat flux must remain below engineering limits, typically of order $10~\mathrm{MW/m^2}$ \citep{stangeby2018basic,pitts2011physics}, to avoid unacceptable erosion or melting of plasma-facing components. With DEMO-class devices expected to operate at even higher power densities, substantially stronger flux reduction in the divertor will be required \citep{najmabadi2006aries,tobita2007slimcs,leonard2018plasma}.

At sufficiently high plasma collisionality, the divertor can enter a cold, highly radiative regime in which the plasma loses energy through radiation and other dissipative processes, while also losing momentum through charge exchange, inelastic collisions, and recombination. These mechanisms reduce the ion flux reaching the target, lower the electron pressure along the field line, and drive the electron temperature near the target to only a few electronvolts \citep{soukhanovskii2017review}. When the target electron temperature falls below roughly $5$~eV \citep{militello2022boundary}, volumetric power and momentum losses become strong enough for the ionization front to move away from the plate, leading to plasma recombination and detachment. This radiative operating regime is commonly referred to as divertor plasma detachment \citep{krasheninnikov2017physics}.

The onset and stability of detachment depend on several control parameters, including the separatrix electron density, the power entering the scrape-off layer (SOL), the impurity seeding rate, and the external gas fueling rate \citep{lipschultz2016sensitivity}. In general, higher upstream density facilitates detachment by increasing collisionality and radiative losses. However, the operational window is constrained: excessive impurity seeding can degrade core performance, promote the formation of X-point multifaceted asymmetric radiation from the edge (MARFE) instabilities, and reduce plasma confinement.

Predictive modeling of detachment requires the self-consistent treatment of plasma transport, atomic physics, radiation, and neutral dynamics. There are two main classes of models for neutral particle transport. The first solves the neutral kinetic equation using Monte Carlo (MC) methods, such as EIRENE \citep{reiter2005eirene}, while the second is based on neutral fluid descriptions \citep{rensink1998comparison,umansky2003modeling,riemann2004navier,hoshino2008benchmarking}. Recent developments and assessments of fluid neutral models have shown that some kinetic effects can be captured by evolving neutral density together with parallel momentum and energy equations \citep{horsten2017development,horsten2017assessment}. In this context, neutral fluid models are attractive because they provide a computationally efficient treatment of ionization, recombination, charge exchange, and neutral transport, while remaining compatible with large-scale edge-plasma simulations.

Nevertheless, detachment remains a multiscale problem that is numerically challenging because it couples stiff source terms, strong spatial gradients, and multiple interacting species. In two-dimensional transport calculations, the main objective is often the steady-state solution. This motivates the development of robust solvers and multi-species models with predictive capability for detached divertor plasmas.

Hermes-3 is an open-source, multi-component plasma code implemented within the BOUT++ framework and has undergone substantial development in recent years \citep{dudson2024hermes}. Written in C++, it provides flexible field operators that facilitate code extension and the incorporation of new physics models. Hermes-3 is designed as a multi-fidelity model, supporting one-, two-, and three-dimensional simulations in both steady-state and time-dependent regimes, including turbulence simulations. The framework is parallelized using MPI, enabling efficient execution on multiple CPUs and making it suitable for large-scale plasma simulations.

In this work, we first present improvements to the nonlinear solver used for two-dimensional axisymmetric transport simulations with neutrals. Solver performance is enhanced through optimization of linear-solver preconditioner parameters and by testing alternative preconditioners, including STRUMPACK \citep{ghysels2017robust}, which provided the most favorable performance for the cases considered here. The improved solver is then used with a 2D axisymmetric fluid model to characterize detached divertor conditions in DIII-D. The individual transport and source terms in the steady-state density and pressure equations are evaluated to identify the dominant physical mechanisms governing particle and energy balance. These developments establish a foundation for future multispecies tokamak simulations with BOUT++/Hermes-3.

The remainder of this paper is organized as follows. Section~\ref{sec:model} introduces the plasma and neutral fluid models implemented in BOUT++/Hermes-3. Section~\ref{sec:NumericalMethod} describes the numerical method and computational setup for 2D transport simulations, with particular emphasis on improvements to the nonlinear solver. Section~\ref{sec:SolverOptimization} evaluates the performance of different preconditioners and solver settings. Section~\ref{sec:Results} applies the model to DIII-D to characterize detached divertor conditions, analyze steady-state particle and energy balances, examine an upstream density scan, and compare parallel heat-flux limiter models. Finally, Section~\ref{sec:Conclusions} summarizes the main findings.


\section{Mathematical Model} \label{sec:model}

Drift-ordered fluid models \citep{braginskii1965transport, mikhailovskii1971transport} are widely used in tokamak edge simulations and encompass a broad range of formulations and levels of complexity. The drift-reduced fluid equations used in Hermes-3 are adapted from the formulations presented in \cite{simakov2003drift,catto2004drift,simakov2004erratum}. Within the drift-reduced approximation, the plasma velocity is decomposed into parallel and perpendicular components. Additional details on the derivation of these models are provided in \cite{militello2022boundary}. 

In the present work, perpendicular transport is modeled through anomalous (turbulence-driven) diffusion, as is commonly done in two-dimensional transport simulations, while the $\mathbf{E}\times\mathbf{B}$, diamagnetic, and polarization drifts are neglected. A detailed description of the plasma and neutral transport equations employed in this work is provided in Appendix B of \cite{dudson2024hermes} and online in the Hermes-3 documentation \citep{hermes3_docs}. For brevity, we summarize here only the conservation equations for density, momentum, and energy for each plasma species. SI units are adopted throughout this work, except for temperature, which is expressed in eV.

The governing equations for the deuterium ion density, $n_{d+}$, parallel velocity, $v_{\parallel,d+} \equiv \mathbf{b}\cdot\mathbf{v}_{d+}$, ion pressure, $p_{d+} = e\,n_{d+}T_{d+}$, and electron pressure, $p_e$, are given by:

\begin{equation}
\label{eq:ion_number_density}
\frac{\partial n_{d+}}{\partial t}
= - \nabla \cdot \left[
\left( \mathbf{b} v_{\parallel,d+} + \mathbf{v}_{\perp,d+} \right) n_{d+}
\right]
+ R_{iz} - R_{rc},
\end{equation}

\begin{multline}
\label{eq:ion_momentum}
\frac{\partial}{\partial t}
\left( m_{d+}\, n_{d+}\, v_{\parallel,d+} \right)
=
- \nabla \cdot
\left[
\left( \mathbf{b} v_{\parallel,d+}
    + \mathbf{v}_{\perp,d+}
\right)
\left( m_{d+}\, n_{d+}\, v_{\parallel,d+} \right)
\right]
- \mathbf{b} \cdot \nabla p_{d+} 
\\
+ \nabla \cdot \left( \eta_{\perp,d+}\, \nabla v_{\parallel,d+} \right) 
+ \sqrt{B}\, \nabla \cdot \left[ \frac{\eta_{d+}}{B} \mathbf{bb}\cdot \nabla \left(\sqrt{B}\,v_{\parallel,d+}\right) \right] + e\, n_{d+}\, E_{\parallel} 
\\
+ R_{cx}\, m_d \left( v_{\parallel,d} - v_{\parallel,d+} \right)
- R_{rc}\, m_d\, v_{\parallel,d+}
+ R_{iz}\, m_d\, v_{\parallel,d},
\end{multline}

\begin{multline}
\label{eq:ion_pressure}
\frac{\partial}{\partial t}\!\left(\frac{3}{2} p_{d+}\right)
=
- \nabla \cdot \!\left[
\left( \frac{5}{2} \mathbf{b} v_{\parallel,d+} + \frac{3}{2}\mathbf{v}_{\perp,d+} \right)
p_{d+}
\right]
+ v_{\parallel,d+}\, \mathbf{b} \cdot \nabla p_{d+} 
\\
+ \nabla \cdot \left( \mathbf{b}\, \kappa_{\parallel,d+}\, \mathbf{b} \cdot \nabla T_{d+} \right)
+ \nabla \cdot \left( \chi_{\perp,d+}\, n_{d+}\, \nabla_{\perp} T_{d+} \right)
- v_{\parallel,d+} \sqrt{B}\, \nabla \cdot \left[ \frac{\eta_{d+}}{B} \mathbf{bb}\cdot \nabla \left(\sqrt{B}\,v_{\parallel,d+}\right) \right]
\\
+ W_{d+,e}
+ R_{iz}\, \frac{3}{2} e\, T_d
- R_{rc}\, \frac{3}{2} e\, T_{d+}
+ R_{cx}\, \frac{3}{2}\, e\, \left( T_d - T_{d+} \right)
+ \frac{1}{2} m_d\, (R_{cx} + R_{iz})\, \left( v_{\parallel,d} - v_{\parallel,d+} \right)^2,
\end{multline}

\begin{multline}
\label{eq:electron_pressure}
\frac{\partial}{\partial t}\!\left(\frac{3}{2} p_{e}\right)
=
- \nabla \cdot \!\left[
\left( \frac{5}{2}\mathbf{b} v_{\parallel,e} + \frac{3}{2}\mathbf{v}_{\perp,d+} \right)
p_{e}
\right]
+ v_{\parallel,e}\, \mathbf{b} \cdot \nabla p_{e} 
\\
+ \nabla \cdot \left( \mathbf{b}\, \kappa_{\parallel,e}\, \mathbf{b} \cdot \nabla T_{e} \right)
+ \nabla \cdot \left( \chi_{\perp,e}\, n_{e}\, \nabla_{\perp} T_{e} \right) 
- E_{iz}
+ E_{rc}
+ W_{e,d+} + W_{e,d}  + P_{\mathrm{rad}},
\end{multline}

\noindent where $\mathbf{b} = \mathbf{B}/B$ is the unit vector along the magnetic field, and the gradient perpendicular to the magnetic field is defined as $\nabla_{\perp} \equiv \nabla - \mathbf{b}\,\mathbf{b}\cdot\nabla$. Particle diffusion across the magnetic field is implemented through the cross-field ion drift velocity $\mathbf{v}_{\perp,d+}$, with diffusion coefficient $D_{\perp}$:

\begin{equation}
\mathbf{v}_{\perp,d+}
= - D_{\perp}\,\frac{1}{n_{d+}}\,
\nabla_{\perp} n_{d+},
\end{equation}

\noindent where $\chi_{\perp,d+}$ and $\chi_{\perp,e}$ are the perpendicular heat diffusivities of deuterium ions and electrons, respectively, and appear in the anomalous cross-field thermal transport terms. The anomalous perpendicular viscosity coefficient, which appears in the deuterium ion momentum equation, is given by $\eta_{\perp,d+} = D_{\perp} m_{d+} n_{d+}$. Parallel Braginskii ion viscosity is also included in the deuterium ion momentum equation, with $\eta_{d+}=1.28,p_{d+}\tau_{d+}$, where $\tau_{d+}$ is the ion collision time. This viscosity is flux limited in the confined region.

The quantities $R_{cx}$, $R_{iz}$, and $R_{rc}$ denote the charge-exchange (CX), ionization (IZ), and recombination (RC) reaction rates, respectively, obtained from the AMJUEL database \citep{reiter2000data}. The quantities $\kappa_{\parallel,d+}$ and $\kappa_{\parallel,e}$ denote the parallel ion and electron thermal conductivities, respectively. Their calculation and the heat-flux-limiting models used in the present work are discussed further in Section~\ref{sec:FluxLimiters}. The term $W_{d+,e}$ represents the collisional transfer of thermal energy from electrons to deuterium ions and is given by:
\begin{equation}
W_{d+,e} = 3 \nu_{d+,e}n_{d+} \frac{m_{d+}}{m_{d+}+m_e} e \left(T_e-T_{d+}\right),
\end{equation}
where $\nu_{d+,e}$ is the ion-electron collision frequency and $m_e$ is the electron mass. The terms $W_{e,d+}$ and $W_{e,d}$ are defined analogously for collisional energy exchange between electrons and ions and between electrons and neutral deuterium, respectively. $m_d$ is the deuterium mass and $e$ is the elementary charge. The factor $e$ converts the temperatures, expressed in eV, to energy units. The terms $E_{iz}$ and $E_{rc}$ correspond to the energy loss and gain associated with ionization and recombination, respectively.

The electron density is determined from quasineutrality, $n_e = n_{d+}$, while the electron parallel velocity is assumed equal to the ion parallel velocity, $v_{\parallel,e} = v_{\parallel,d+}$, corresponding to the assumption of zero parallel current. The parallel electric field, $E_{\parallel} \equiv \mathbf{b}\cdot\mathbf{E}$, is calculated from the electron force balance. The resulting electric field is then applied as a force on the ion species.

\begin{equation}
e\, n_{e}\, E_{\parallel}
=
-\, \mathbf{b}\cdot\nabla p_{e}
- \nabla\cdot\!\left[
\mathbf{v}_{\perp,d+}\,
\left( m_{e}\, n_{e}\, v_{\parallel,e} \right)
\right].
\end{equation}

\subsection{Neutral Fluid Model} \label{sec:neutral_model}

The neutral fluid model implemented in Hermes-3 is similar to the formulations of \cite{horsten2017development,horsten2017assessment}. The governing equations for the neutral deuterium density, $n_d$, parallel velocity, $v_{\parallel,d}$, and pressure, $p_d = e\,n_d T_d$, are given by:

\begin{equation}
\frac{\partial n_{d}}{\partial t}
=
-\,\nabla \cdot
\left[
\left( \mathbf{b}\, v_{\parallel,d} + \mathbf{v}_{\perp,d} \right)
n_{d}
\right]
- R_{iz}
+ R_{rc},
\end{equation}

\begin{multline}
\frac{\partial}{\partial t}
\left( m_{d}\, n_{d}\, v_{\parallel,d} \right)
=
-\,\nabla \cdot
\left[
\left( \mathbf{b} v_{\parallel,d}
      + \mathbf{v}_{\perp,d}
\right)
\left( m_{d}\, n_{d}\, v_{\parallel,d} \right)
\right]
-\, \mathbf{b} \cdot \nabla p_{d}
+ \nabla \cdot \left( \eta_{d}\, \nabla v_{\parallel,d} \right) \\
- R_{cx}\, m_{d} \left( v_{\parallel,d} - v_{\parallel,d+} \right)
+ R_{rc}\, m_{d}\, v_{\parallel,d+}
- R_{iz}\, m_{d}\, v_{\parallel,d},
\end{multline}

\begin{multline}
\frac{\partial}{\partial t}\!\left(\frac{3}{2} p_{d}\right)
=
-\,\nabla \cdot \!\left[
\left( \mathbf{b} v_{\parallel,d}
     + \mathbf{v}_{\perp,d}
\right)
\left( \frac{5}{2} p_{d} \right)
\right]
+ v_{\parallel,d}\, \mathbf{b}\cdot\nabla p_{d}
+ \nabla \cdot \left( \kappa_{d}\, \nabla T_{d} \right)
-\, v_{\parallel,d}\, \nabla\cdot\left( \eta_{d}\, \nabla v_{\parallel,d} \right)
\\
+ W_{d,e} 
- R_{iz}\, \frac{3}{2} e\, T_{d}
+ R_{rc}\, \frac{3}{2} e\, T_{d+}
+ R_{cx}\, \frac{3}{2}\, e\, \left( T_{d+} - T_d  \right)
+ \frac{1}{2} m_{d}\,\left( R_{cx} + R_{rc} \right) \left( v_{\parallel,d} - v_{\parallel,d+} \right)^{2}.
\end{multline}

Neutral particles are not magnetically confined and can therefore migrate upstream across magnetic flux surfaces. The perpendicular velocity of neutral atoms, $\mathbf{v}_{\perp,d}$, is obtained by neglecting inertial terms in the perpendicular momentum equation and balancing the friction force against the pressure gradient \citep{horsten2017development,van2020implementation}:

\begin{equation}
\mathbf{v}_{\perp,d}
=
- D_{d}\,\frac{1}{p_{d}}\,
\nabla_{\perp} p_{d},
\end{equation}

\begin{equation}
D_{d} = \frac{v_{\text{th}}^{2}}{\nu_{\text{tot}}},
\end{equation}

\noindent where $v_{\mathrm{th}}$ is the thermal velocity of neutral atoms and $\nu_{\mathrm{tot}}$ is the total collision frequency with other species, including charge exchange, ionization, and neutral--neutral collisions. The corresponding transport coefficients are given by:

\begin{equation}
\kappa_{d}
=
\frac{5}{2}\,D_{d}\,n_{d},
\end{equation}

\begin{equation}
\eta_{d}
=
\frac{2}{5}\,m_{d}\,\kappa_{d}.
\end{equation}

To limit neutral transport in regions where the neutral mean free path becomes comparable to the local gradient scale length, the following flux limiter is applied \citep{umansky2003modeling}. The limiter constrains the fluxes to values comparable to the local free-streaming limit:

\begin{equation}
\Gamma_{\mathrm{fl}}
=
\Gamma \left(
1 + \left(
\frac{\Gamma}{\alpha\,\Gamma_{\mathrm{lim}}}
\right)^{\gamma_{\mathrm{fl}}}
\right)^{-1/\gamma_{\mathrm{fl}}}.
\end{equation}

\noindent Here, $\Gamma_{\mathrm{lim}}$ denotes the limiting flux, while typical parameter values are $\alpha = 1$ and $\gamma_{\mathrm{fl}} = 1$. More information can be found at \cite{hermes3_docs}.


\section{Numerical method} \label{sec:NumericalMethod}

Hermes-3 is a fluid plasma model implemented within the BOUT++ framework for simulations of tokamak edge  plasmas. It solves a set of drift-reduced fluid equations using a finite-difference discretization on field-aligned grids, which enables efficient treatment of the strong anisotropy along magnetic field lines. It has also been applied to 1D \citep{dudson2019role,body2024scalings}, 2D \citep{dudson2024hermes}, and 3D \citep{dudson2025validation,chang2025hermes} tokamak simulations, and more recently to stellarator configurations \citep{tork2025bout++}.

In this section, we describe the numerical methods used in Hermes-3 that are relevant to the results presented in Section~\ref{sec:Results}. This description is not intended to be exhaustive, since the spatial discretization has already been documented elsewhere \citep{dudson2024hermes}; instead, the focus here is on the time-integration solver. In Hermes-3, the system of partial differential equations is solved using the method of lines, in which the spatial and temporal dimensions are treated separately, so that the time integrator advances a coupled set of ordinary differential equations (ODEs). Solver performance is improved through optimization of the preconditioner parameters, as discussed in more detail below.

Hermes-3 employs conservative finite-difference operators for the spatial discretization of equations written in terms of fluxes through cell faces. All quantities are cell-centered, and the advection operators are expressed in terms of intercell fluxes evaluated at cell faces. These operators are implemented on grids that are orthogonal to the magnetic field. They are second-order accurate and use monotonized-central (MC) slope limiters \citep{van1977towards} to improve numerical stability and to suppress checkerboard modes that can arise in non-staggered formulations in the absence of dissipation. Additional details of the method are given in Appendix A of \cite{dudson2024hermes}. Along the magnetic field direction, a Lax flux is used to stabilize grid-scale oscillations. Because the present work focuses on two-dimensional transport simulations without drifts, acoustic waves propagating at the sound speed are taken to be the fastest waves in the system for the Lax-flux terms. A more detailed description of the operators is available online in the BOUT++ documentation \citep{boutpp_docs}.

BOUT++ provides two main time-integration solvers, both of which are available in Hermes-3 and are suited to different classes of problems. CVODE, from SUNDIALS \citep{hindmarsh2005sundials}, is a variable-order time integrator that employs matrix-free Jacobian--vector products. In BOUT++, CVODE has also been used  in combination with physics-based preconditioners \citep{chacon2002implicit,chacon2008optimal}. It is mainly used for three-dimensional simulations, such as studies of edge-localized-mode (ELM) crashes \citep{xu2010nonlinear}, tokamak edge electromagnetic turbulence \citep{zhu2023electromagnetic,umansky2024simulations, dudson2025validation}, and SOL turbulence and filament transport \citep{omotani2025storm}, where accurate temporal evolution is essential.

In contrast, the solver employed in this work is tailored to two-dimensional transport simulations, for which the primary objective is to obtain steady-state solutions. Inspired by the approach used in UEDGE~\citep{rognlien1992fully}, it is based on the PETSc SNES nonlinear solver framework and uses a custom implementation of a rearranged backward-Euler method, which provides a robust implicit pseudo-time integration strategy. At each nonlinear iteration the Jacobian is approximated using finite differences, with PETSc's Jacobian coloring \citep{balay2019petsc} used to evaluate multiple perturbations simultaneously and reduce computational cost. The resulting linear systems are solved using GMRES with incomplete LU (ILU) preconditioning. The first application of this solver approach within BOUT++ was reported in \cite{dudson2024hermes}. Its performance and robustness are further improved through the optimization of solver parameters, as discussed in Section~\ref{sec:SolverOptimization}.

Time-step adaptivity is used to improve solver efficiency: If the timestep is too large then the nonlinear iterations will likely not converge or will take too many iterations; A timestep that is too small will make little progress towards the steady state solution. A timestep
controller was implemented to dynamically adjust the time step based on the nonlinear iteration count. Specifically, the time-step multiplication factor for step $k$ is:
\begin{equation}
g_k = 1.1 
\left(\frac{N_{k-1}}{N_k}\right)^{k_P}
\left(\frac{N_{\mathrm{target}}}{N_k}\right)^{k_I}
\left(\frac{N_{k-1}^{2}}{N_k N_{k-2}}\right)^{k_D},
\label{eq:pid_factor}
\end{equation}
where $N_k$ is the nonlinear iteration count for the current step, $N_{k-1}$ and $N_{k-2}$ are the iteration counts for the two preceding steps, and $N_{\mathrm{target}}=5$ is the target iteration count. The parameters $k_P$, $k_I$, and $k_D$ are user-configurable and are set to 0.25, 0.65, and 0.15, respectively. The prefactor 1.1 ensures that the time step continues to increase slightly when the nonlinear iteration count remains at the target value.  The simulation is initialized with small timesteps, which then increase aggressively as the solution approaches a steady state. During each nonlinear solve, the Jacobian is fixed and is recomputed every seven iterations or following a nonlinear convergence failure.

\subsection{Computational setup and numerical details} \label{sec:ComputationalSetUp}

The computational mesh is aligned with magnetic flux surfaces that span the separatrix. The magnetic equilibrium is obtained from a kinetic-EFIT reconstruction \citep{lao1990equilibrium} of DIII-D shot \#174287 at 4000~ms, corresponding to a large Type-I ELM regime, and is provided in GEQDSK format. The magnetic field is assumed to be time-independent in these 2D simulations.

The grid is generated using the Hypnotoad tool \citep{omotani2019hypnotoad}. It is structured and orthogonal in the poloidal planes. The domain consists of a closed-flux-surface region inside the separatrix, where magnetic field lines are confined, and an open-field-line region outside the separatrix, known as the scrape-off layer (SOL), where field lines intersect material surfaces. The private-flux region (PFR), located below the X-point, is also included in the open-field-line region.

\begin{figure}[!ht]
\centering
\includegraphics[width=0.49\linewidth]{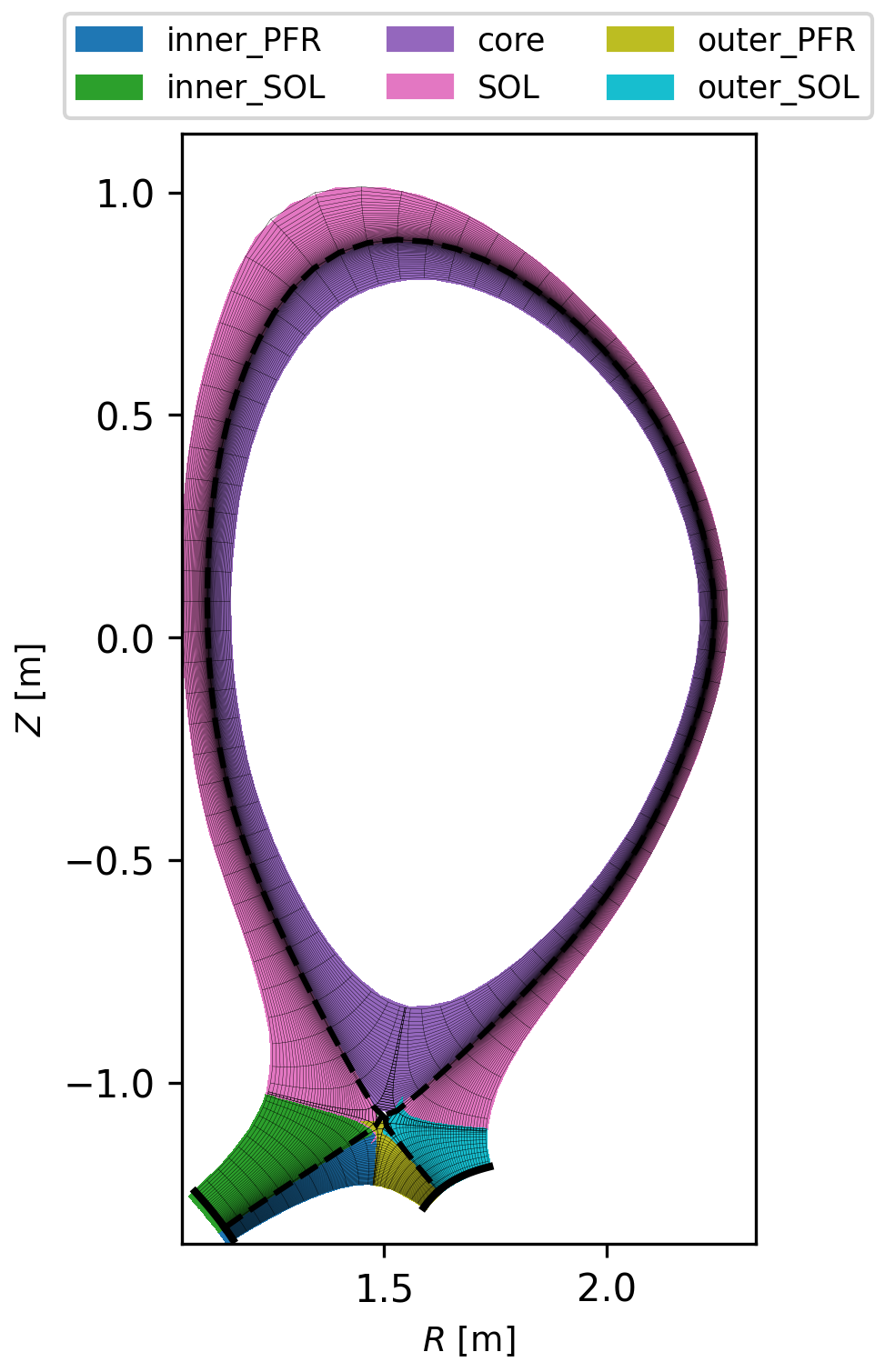}
\includegraphics[width=0.49\linewidth]{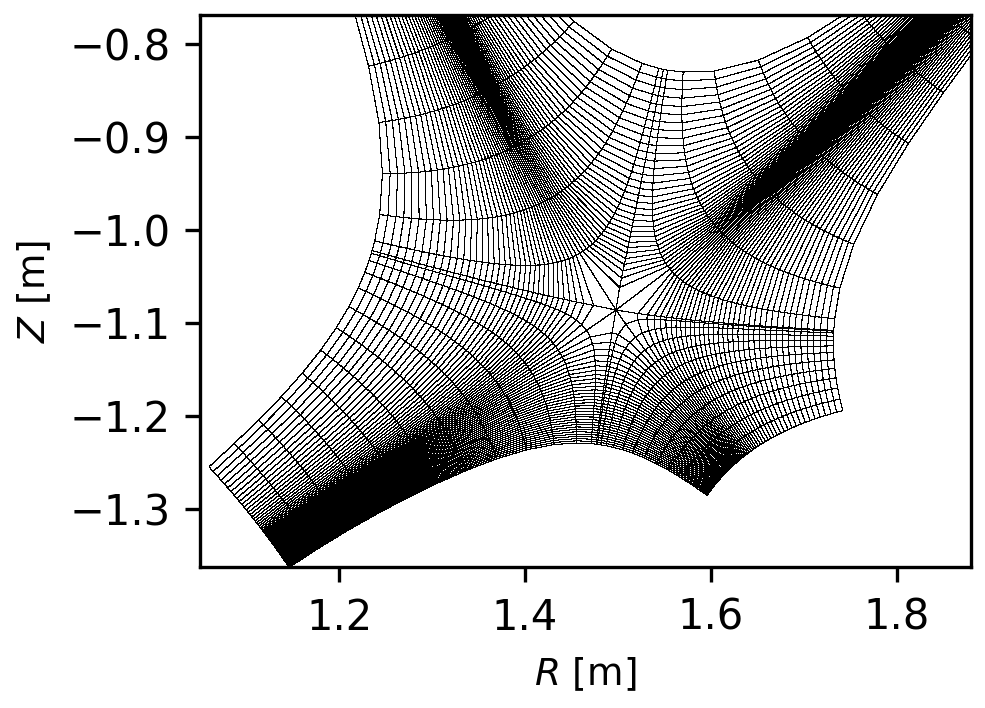}
\caption{Two-dimensional grid used in Hermes-3 for the DIII-D geometry corresponding to shot $\#174287$ at 4000~ms, in a large Type-I ELM regime. The left panel shows the six logically rectangular regions of the grid, while the right panel shows a magnified view of the X-point and divertor regions. The target plates are assumed to lie at the ends of the orthogonal grid. The grid resolution is $64 \times 96$ in the radial and poloidal directions, respectively, with 12 poloidal planes in each divertor leg. The grid is refined in the vicinity of the separatrix.}
\label{fig:Grid_zoom}
\end{figure}

For the single-null X-point geometry considered here, the grid is divided into six logically rectangular regions: the lower inner private-flux region, the lower inner scrape-off-layer region, the core region, the SOL region, the lower outer private-flux region, and the lower outer SOL region as shown in Figure~\ref{fig:Grid_zoom} (left). In the radial ($x$) direction, the boundaries lie on magnetic flux surfaces and do not extend to the device walls. The target plates are represented by the ends of the orthogonal grid rather than by the actual geometry of the plasma-facing components.

A range of computational domains and grid resolutions is employed in the present work. The sensitivity of the results to grid resolution is discussed separately in Section~\ref{sec:GridResolution}. Unless otherwise stated, the simulations presented in Section~\ref{sec:Results} use a grid resolution of $64 \times 96$ in the radial and poloidal directions, respectively, with 12 poloidal planes in each divertor leg. The computational domain covers the range $\psi_{\mathrm{n}} = 0.9$--$1.1$, where $\psi_{\mathrm{n}}$ is the normalized poloidal magnetic flux, defined as $\psi_{\mathrm{n}} = (\psi - \psi_{\mathrm{axis}})/(\psi_{\mathrm{separatrix}} - \psi_{\mathrm{axis}})$, with $\psi$ denoting the poloidal magnetic flux. A portion of the grid focusing on the X-point and divertor regions is shown in Figure~\ref{fig:Grid_zoom} (right). The grid is refined radially in the vicinity of the separatrix, while linear spacing is used in the poloidal ($y$) direction. 

At the innermost flux surface of the computational domain, corresponding to the core boundary, Dirichlet (fixed-value) boundary conditions are applied to the plasma density and pressures. Specifically, the deuterium ion density is set to $N_{d+} = 5.56 \times 10^{19}~\mathrm{m^{-3}}$, while the electron and ion temperatures are fixed at $T_e = 391.3~\mathrm{eV}$ and $T_{d+} = 477.4~\mathrm{eV}$, respectively. The plasma pressure values used in the code are obtained from these quantities. These boundary values are taken from fits to experimental outer-midplane profiles provided through the kinetic-EFIT reconstruction.  Because Dirichlet conditions are imposed at the core boundary, this boundary acts as a source of particles and heat. Hermes-3 also allows the core boundary to be specified in terms of input power, but that option is not used here. For the neutral density, $N_d$, and neutral pressure, $P_d$, Neumann (zero-gradient) boundary conditions are applied at the core.

At the outer SOL boundary, an exponential-decay boundary condition is imposed, with a prescribed decay length of 3~cm for the plasma density and pressures. This condition represents cross-field losses in the far scrape-off layer. The resulting ion fluxes are recycled into neutrals at the outer radial domain edge.

In contrast to earlier Hermes-3 simulations, Neumann (zero-gradient) boundary conditions are imposed on the parallel velocities (or momenta) of all species, including neutrals, at the radial boundaries. This treatment reduces steep radial gradients near the radial boundaries that can adversely affect solver convergence. Similar boundary conditions for the parallel velocities were employed in \citep{rognlien1999two}. 

At the boundaries of the domain in the $y$ direction, pseudo-periodic boundary conditions are applied within the closed-field-line region, while a simple Bohm sheath boundary condition \citep{stangeby1995bohm} is imposed at the divertor targets along open field lines. At the targets, the ion parallel velocity satisfies the Bohm criterion, reaching or exceeding the local sound speed. The parallel heat flux to the targets is given by $q_{\parallel,e,d+} = \gamma_{e,d+} n_e T_{e,d+} c_s$, where $\gamma_e = 4.0$ and $\gamma_{d+} = 3.5$ are the sheath heat transmission coefficients for electrons and ions, respectively. The sound speed at the sheath entrance is defined as $c_s = \sqrt{e\,(T_{d+} + T_e)/m_i}$. For the neutral species, a free boundary condition is applied to the neutral density $N_n$, a Neumann boundary condition is applied to the neutral pressure $P_n$, and a Dirichlet condition $V_n = 0$ is imposed on the neutral velocity.

Deuterium ions incident on the divertor targets and outer radial boundary are recycled and re-emitted as neutral atoms, with a recycling coefficient of 0.99. A small neutral source is imposed near the top of the domain, and neutral pumping with an albedo of 0.99 is applied in the PFR, though their effect is negligible for the cases considered. Physical and chemical sputtering of carbon impurities at the walls and target plates has been neglected. Molecular species are also not included, and implementation of these models is left for future work.

The plasma state variables are initialized using fits to experimental outer-midplane profiles provided through the kinetic-EFIT reconstruction. These profiles provide values of $N_e$, $T_e$, and $T_{d+}$ inside the separatrix, while constant values are prescribed in the scrape-off layer, where experimental data are unavailable. The resulting profiles are mapped uniformly along the poloidal direction; that is, no initial poloidal variation is assumed. Such profiles are commonly used in BOUT++ six-field simulations for linear and nonlinear ELM studies~\citep{li2025exploring}. Because the present simulations are evolved to steady state, the final solution is typically not affected by the choice of initial conditions, except in cases where there are bifurcated solutions~\cite{wigram2019arc}. The initial ion parallel velocity $V_{d+}$ is set to zero. The neutral density $N_d$ is initialized to $10^{17}\,\mathrm{m^{-3}}$ throughout the domain, and the initial neutral temperature $T_d$ is set to $40\,\mathrm{eV}$.

The equations in Hermes-3 are solved in normalized form using the reference quantities $N_0 = 10^{19}\,\mathrm{m^{-3}}$, $B_0 = 1\,\mathrm{T}$, and $T_0 = 10\,\mathrm{eV}$. The corresponding derived scales are the sound speed $c_{s0} = \sqrt{e T_0 / m_p}$, the ion cyclotron frequency $\Omega_{ci} = e B_0 / m_p$, and the ion Larmor radius $\rho_{s0} = c_{s0} / \Omega_{ci}$, where $e$ is the elementary charge and $m_p$ is the proton mass.

Hermes-3 is publicly available on GitHub under the GPL-3 open-source license. The numerical results presented in this work were obtained using Hermes-3 commit 2f8a496 and BOUT++ commit 6f01052. The full commit hashes, associated \texttt{BOUT.inp} input files, and \texttt{options.yml} files containing the grid-generation parameters are included in the supplementary material. All simulations were performed using PETSc version 3.24.4.


\subsection{Steady-state solution and solver optimizations} \label{sec:SolverOptimization}

\begin{table}[t]
\centering
\caption{Preconditioner and sparse-solver settings used in the steady-state Hermes-3 simulations. Default values are given in parentheses. The complete set of solver parameters is provided in the corresponding \texttt{BOUT.inp} input files included in the supplementary material.}
\label{tab:preconditioner_settings}
\footnotesize
\begin{tabular}{>{\raggedright\arraybackslash}p{2.4cm} >{\raggedright\arraybackslash}p{4.2cm} >{\raggedright\arraybackslash}p{2.8cm} >{\raggedright\arraybackslash}p{4.0cm}}
\toprule
Method & Parameter & Value & Description / Comment\\
\midrule
\textbf{\HYPRE{} ILU(k)}
& \path{pc_hypre_ilu_type}
& Block-Jacobi-ILUk
& Level-based ILU factorization. \\
& \path{pc_hypre_ilu_level}
& 1 (0)
& Fill level. \\
& \path{pc_hypre_ilu_local_reordering}
& true (false)
&  \\
& \path{pc_hypre_ilu_tri_solve}
& true (false)
&  \\
\midrule
\textbf{\HYPRE{} ILU(t)}
& \path{pc_hypre_ilu_type}
& Block-Jacobi-ILUT
& Threshold-based ILU factorization. \\
& \path{pc_hypre_ilu_drop_threshold}
& $3\times10^{-4}$ (0.0)
&  \\
& \path{pc_hypre_ilu_max_nnz_per_row}
& 100 (20)
& Maximum number of nonzeros per row. \\
& \path{pc_hypre_ilu_local_reordering}
& true (false)
&  \\
& \path{pc_hypre_ilu_tri_solve}
& true (false)
&  \\
\midrule
\textbf{STRUMPACK ILU}
& \path{pc_factor_mat_solver_type}
& strumpack
& Used as a preconditioner. \\
& \path{pc_type}
& ilu
& ILU preconditioning with low-rank compression. \\
& \path{mat_strumpack_compression}
& HSS (BLR) 
& Negligible differences between BLR and HSS. \\
\bottomrule
\end{tabular}
\end{table}

For axisymmetric tokamak transport simulations, the primary objective is typically to obtain a steady-state solution. Once converged, such solutions enable efficient parametric studies without recomputing the full transient evolution, and can also serve as background states for subsequent three-dimensional turbulence simulations.

PETSc provides a wide range of solver and preconditioning options, allowing systematic exploration of solver strategies. However, default settings are not always optimal for the stiff nonlinear systems encountered in edge plasma simulations. In particular, extending the computational domain further into the core introduces higher temperatures and increased stiffness, making convergence more challenging. In this work, solver performance is improved by tuning preconditioner parameters and comparing several PETSc preconditioning strategies, including \HYPRE{} ILU(k) \citep{hysom1999efficient}, \HYPRE{} ILU(t) \citep{falgout2002hypre}, and STRUMPACK \citep{ghysels2017robust}.

Table~\ref{tab:preconditioner_settings} summarizes the preconditioner and sparse-solver configurations used in the steady-state calculations, along with their default values for reference. For \HYPRE{}, both level-based ILU(k) and threshold-based ILU(t) factorizations were explored. For STRUMPACK, an ILU-based preconditioner with low-rank compression was employed. The Hierarchically Semi-Separable (HSS) compression was used, with only minor differences observed compared to the default Block Low Rank (BLR) compression. The complete set of solver parameters is provided in the corresponding \texttt{BOUT.inp} input files included in the supplementary material.

\begin{figure}[!ht]
\centering
\includegraphics[width=0.49\linewidth]{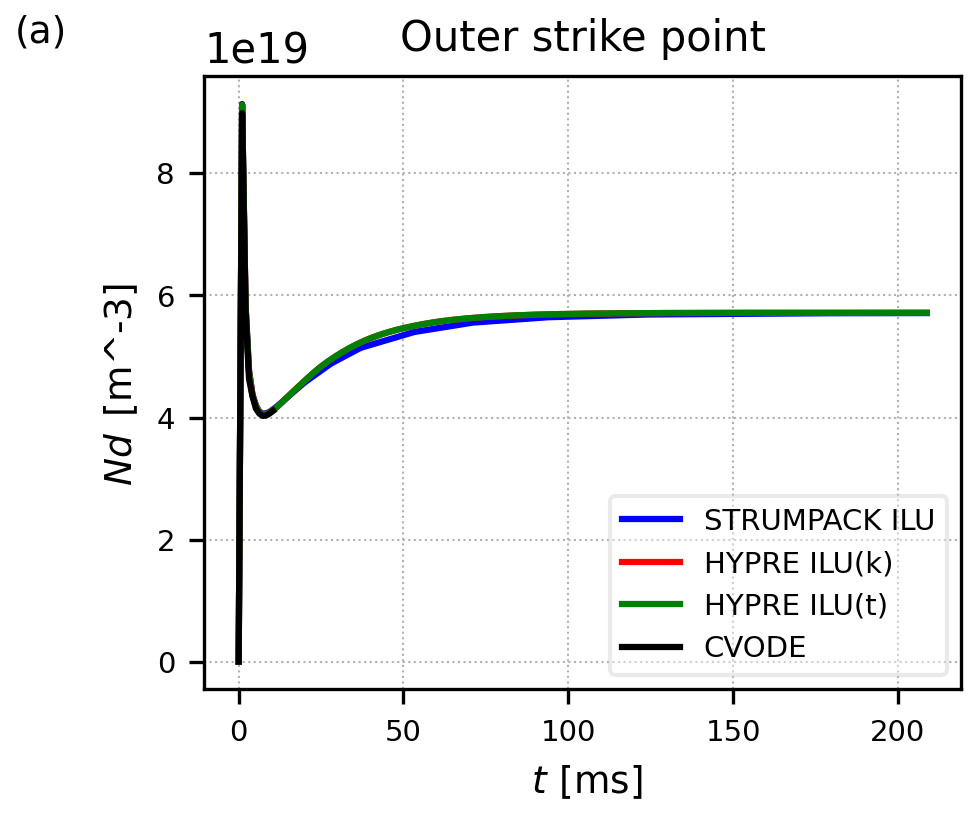}
\includegraphics[width=0.49\linewidth]{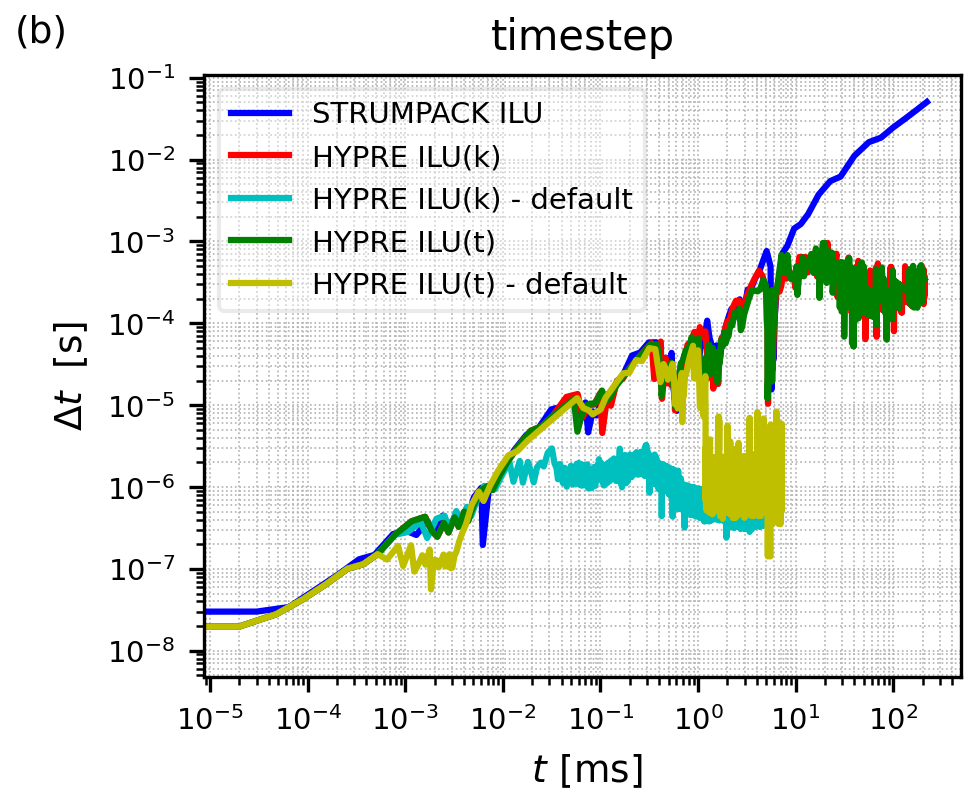}
\includegraphics[width=0.49\linewidth]{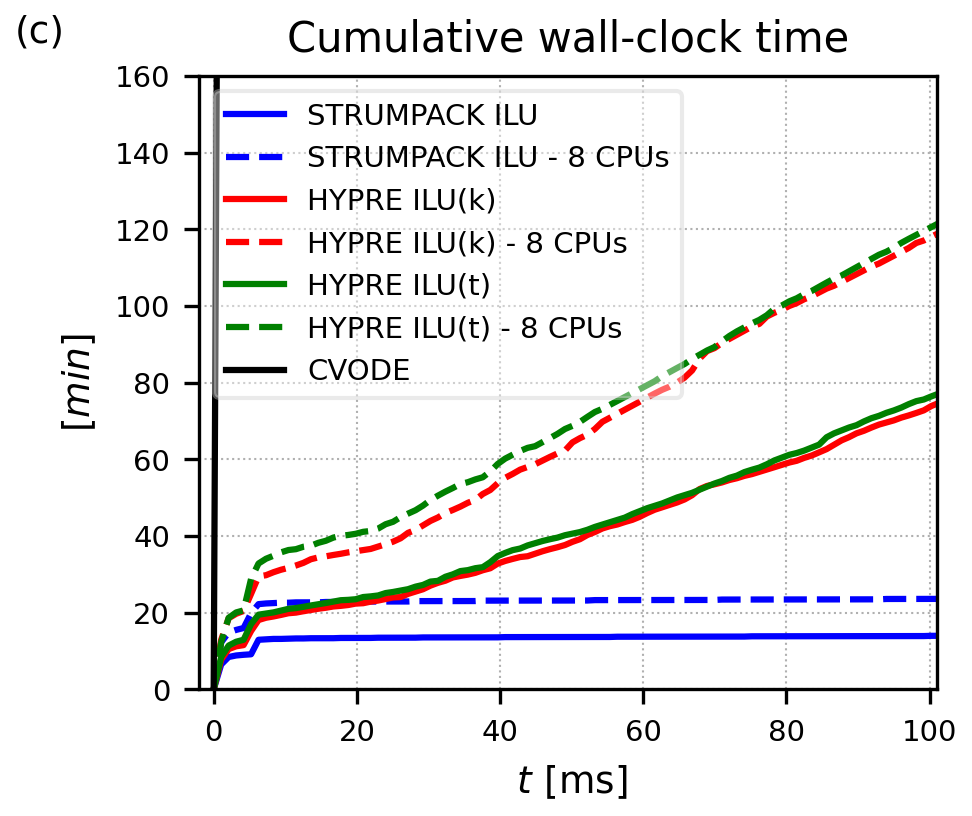}
\includegraphics[width=0.49\linewidth]{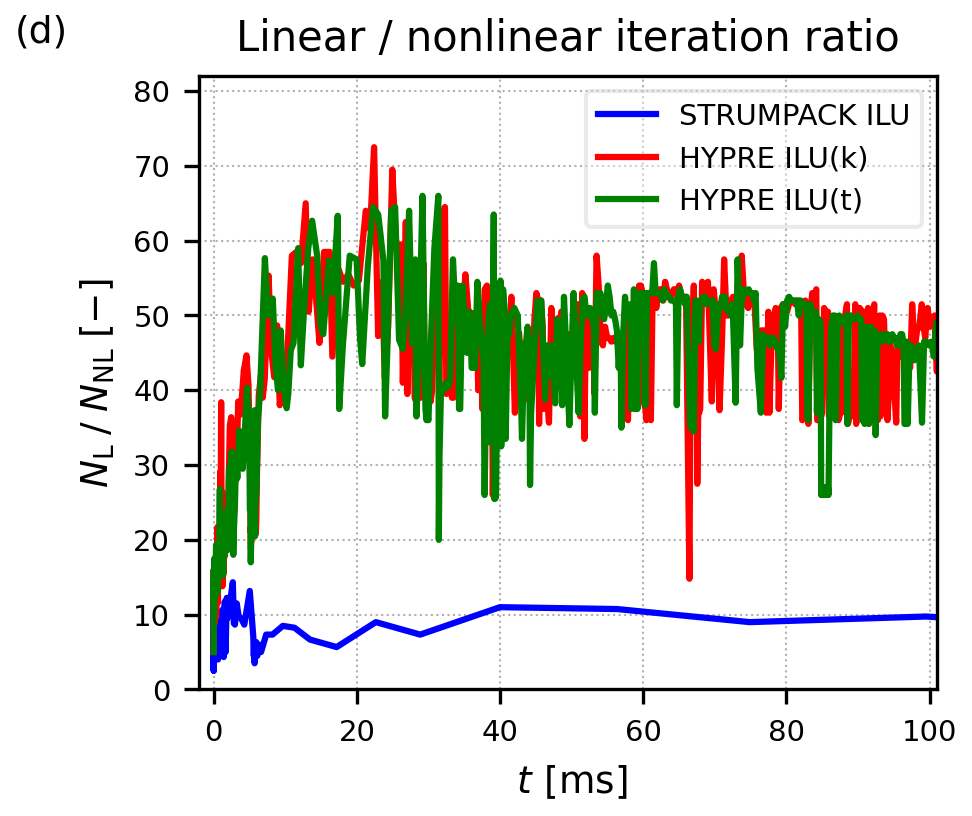}
\caption{Performance of the steady-state solver for the case described in Section~\ref{sec:ComputationalSetUp}: (\textit{a}) time evolution of the neutral density, $N_d$, at the outer divertor target; (\textit{b}) time-step size; (\textit{c}) cumulative wall-clock time; and (\textit{d}) ratio of linear to nonlinear iterations. Simulations were carried out on 16 CPUs, with additional results on 8 CPUs shown in panel (\textit{c}). Solver parameters are listed in Table~\ref{tab:preconditioner_settings}. The horizontal axis in all panels represents the physical (simulation) time in ms rather than wall-clock time.}
\label{fig:performance_mid}
\end{figure}

Figure~\ref{fig:performance_mid}(a) shows the time evolution of the neutral density, $N_d$, at the outer target for the three solver configurations: STRUMPACK used as an ILU preconditioner with low-rank compression, \HYPRE{} ILU(k), and \HYPRE{} ILU(t). Results obtained with the CVODE solver, using physics-based preconditioning applied to the plasma variables, are also shown in the figure. The horizontal axis represents the physical (simulation) time in ms. 
The computational setup of the simulations is described in Section~\ref{sec:ComputationalSetUp}, while a $50 \times 64$ grid is employed for the solver performance study presented here.
For all cases, the absolute and relative tolerances were set to $10^{-12}$ and $10^{-6}$, respectively, while the PETSc SNES convergence tolerance (stol) was not used (i.e. it was set to zero).

In all cases, the solution approaches steady state at approximately 100~ms, and the simulations were continued beyond 200~ms to verify convergence. Only the neutral density at the target is shown in Figure~\ref{fig:performance_mid}(a), as the particle-throughput timescale is typically longer than that required for the power balance to reach steady state \citep{dudson2024hermes}. In addition, convergence in the divertor region generally occurs more slowly than upstream in the core. The remaining plasma and neutral variables were also observed to converge to steady state, and all solver configurations approached the same steady-state solution. By the end of the simulations, the global residual had decreased to approximately $10^{-11}$ for all cases.

Figure~\ref{fig:performance_mid}(b) shows the time-step size $\Delta t$ (s) for the different solver configurations, including \HYPRE{} ILU(k) and ILU(t) with both default and optimized settings, as well as STRUMPACK. In all cases, the simulation is initialized with a small time step ($10^{-8}$ s), which increases progressively through the adaptive time-stepping strategy described in Section~\ref{sec:NumericalMethod}. 

For the \HYPRE{} configurations with default settings, the time step quickly plateaus at very small values. In particular, the ILU(k) case stagnates at $\Delta t \sim 10^{-6}$ s after $\sim 0.01$ ms of simulation time and subsequently decreases to values near $\Delta t \sim 5 \times 10^{-7}$~s, while the ILU(t) case reaches a slightly larger plateau of $\Delta t \sim 2 \times 10^{-6}$ s after $\sim 1$ ms. These small time steps severely limit progress toward steady state, as reaching it would require many days of computation. Time steps larger than $\sim 10^{-6}$ s are generally desirable to realize the computational advantage of fully implicit solvers over explicit methods \citep{schneider2007challenges}.
Using the optimized \HYPRE{} parameters listed in Table~\ref{tab:preconditioner_settings} significantly improves solver performance. For both the ILU(k) and ILU(t) configurations, the time step increases to approximately $\Delta t \sim 2 \times 10^{-4}$~s within the first $\sim 5$~ms of simulation time and subsequently plateaus. This corresponds to an increase of approximately two orders of magnitude compared with the default settings, allowing the simulations to advance to physically relevant timescales.

The STRUMPACK ILU preconditioner exhibits the best performance among the cases considered. In this configuration, the time step continues to increase as the solution approaches steady state, reaching values as large as $\sim 10^{-1}$ s. This improved performance is also reflected in Figure~\ref{fig:performance_mid}(c), which shows the cumulative wall-clock time for each case. The STRUMPACK run requires approximately 16 minutes on 16 AMD EPYC 7763 CPU cores to reach 100 ms of simulation time, compared to about 75 minutes for the \HYPRE{}-based cases. For reference, Figure~\ref{fig:performance_mid}(c) also includes results from the CVODE solver, which required approximately 24 hours to reach 10 ms of simulation time. The almost $5 \times$ speedup achieved with STRUMPACK over the \HYPRE{}-based cases can be attributed to a significantly lower linear-to-nonlinear iteration ratio. As shown in Figure~\ref{fig:performance_mid}(d), the \HYPRE{}-based configurations require roughly 50 linear iterations per nonlinear iteration, whereas the STRUMPACK ILU case requires only about 10.

In Figure~\ref{fig:performance_mid}(c), we also present results obtained with 8 CPUs and compare them with the corresponding 16-CPU cases. The 8-CPU runs require approximately twice the wall-clock time for the STRUMPACK ILU case, suggesting near-linear scaling. A more rigorous assessment of strong scaling would be desirable; however, this was not pursued here due to limitations in the current BOUT++ domain decomposition.

The performance of the \HYPRE{}-based configurations could likely be further improved through additional tuning of the linear GMRES solver. The problem sizes considered here are representative of 2D transport calculations; however, they may increase significantly when additional ion species are included. As a result, the relative performance of the different preconditioners is problem-dependent, and no single approach is expected to be universally optimal. Nevertheless, the performance evaluation and methodology presented here provide useful guidance for future applications and similar test cases.


\subsection{Parallel heat-conduction discretization and grid resolution} \label{sec:GridResolution}


The discretization of the parallel heat-conduction operator was found to have an important effect on nonlinear solver robustness at higher grid resolutions. For a field $f$ and parallel conductivity $K$, Hermes-3 evaluates $\nabla_{\parallel}\cdot\left(K\nabla_{\parallel}f\right)$ using a conservative finite-difference discretization. Denoting the flux through the face between cells $j$ and $j+1$ by $F_{j+1/2}$, the operator in cell $j$ is calculated as
\begin{equation}
\left[\nabla_{\parallel}\cdot\left(K\nabla_{\parallel}f\right)\right]_j
=
\frac{F_{j+1/2}-F_{j-1/2}}
{J_j\Delta y_j}.
\label{eq:parallel_conduction_fv}
\end{equation}
In the original discretization, conductivity $K$ and Jacobian $J$ are independently averaged from cell centers to the cell faces. The flux through the $j+1/2$ face is
\begin{equation}
F_{j+1/2}^{\mathrm{Original}} = \frac{ \overline{K}_{j+1/2} \, \overline{J}_{j+1/2} }{ \overline{g_{yy}}_{j+1/2}} \, \frac{f_{j+1}-f_j}{\overline{\Delta y}_{j+1/2}},
\end{equation}
where the overbar denotes an arithmetic average between the two neighboring cell centers, for example,
\begin{equation}
\overline{K}_{j+1/2} = \frac{K_j+K_{j+1}}{2}, \qquad \overline{J}_{j+1/2} = \frac{J_j+J_{j+1}}{2},
\end{equation}
and analogous arithmetic averages used for metric component $g_{yy}$ and cell spacing parallel to the magnetic field $\Delta y$. 
The alternative \texttt{ProductJK} discretization instead forms the product $KJ$ at the cell centers and then averages this product to the cell face,
\begin{equation}
F_{j+1/2}^{\mathrm{ProductJK}} = \frac{\overline{KJ}_{j+1/2}}{\overline{g_{yy}}_{j+1/2}} \, \frac{f_{j+1}-f_j}{\overline{\Delta y}_{j+1/2}},
\end{equation}
where
\begin{equation}
\overline{KJ}_{j+1/2} = \frac{K_jJ_j+K_{j+1}J_{j+1}}{2}.
\end{equation}

Although the two formulations differ only in how $K$ and $J$ are interpolated to the cell face, this distinction can become important in strongly compressed regions of the mesh. Small-volume cells can occur near mesh seams in both the core and divertor regions. Since the parallel thermal conductivity can be very large, independently averaging $K$ and $J$ can make the conductive flux particularly sensitive to conductivity in small cells. The \texttt{ProductJK} formulation reduces this sensitivity by averaging the locally paired product $KJ$.

The \texttt{ProductJK} method improved convergence of the higher-resolution simulations, enabling them to more reliably reach steady-state solutions for most of the cases considered here. Comparisons with the original formulation showed negligible differences in the converged plasma profiles. The \texttt{ProductJK} discretization is therefore used for all simulations presented in Section~\ref{sec:Results}.


\begin{figure}[!ht]
\centering
\includegraphics[width=0.49\linewidth]{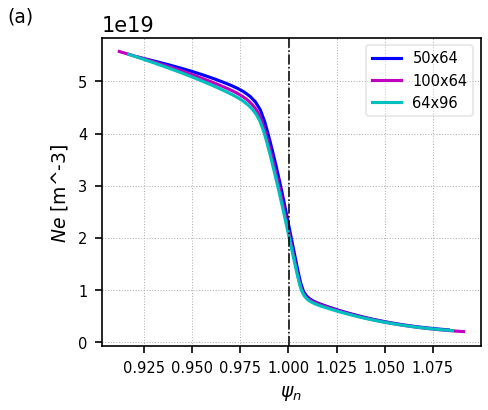}
\includegraphics[width=0.49\linewidth]{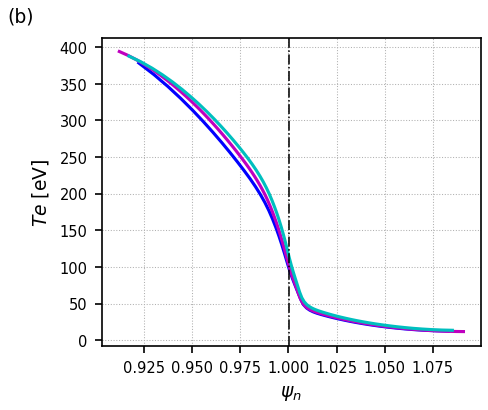}
\includegraphics[width=0.49\linewidth]{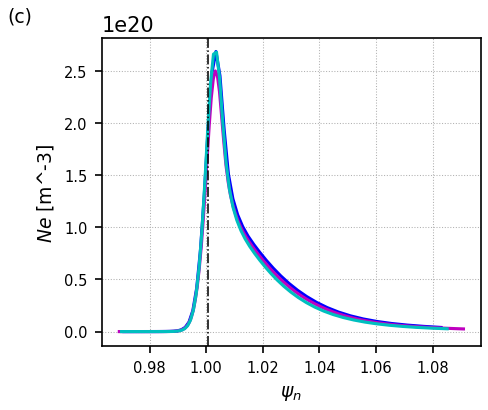}
\includegraphics[width=0.49\linewidth]{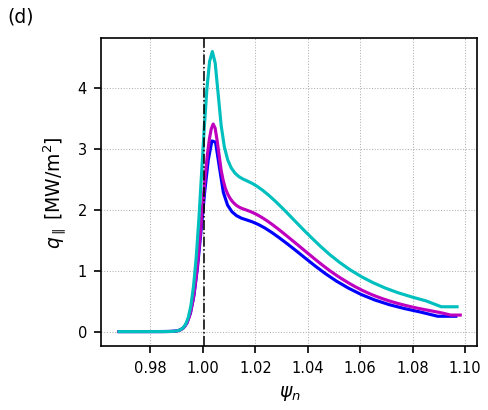}
\caption{Grid-resolution comparison of the steady-state profiles of (a) electron density, $N_e$, at the outer midplane, (b) electron temperature, $T_e$, at the outer midplane, (c) electron density at the outer target, and (d) total parallel heat flux at the outer target. The profiles are shown for $(N_x,N_y)=(50,64)$, $(100,64)$, and $(64,96)$, where $x$ and $y$ denote the radial and poloidal directions, respectively.}
\label{fig:grid_resolution}
\end{figure}

The robustness of the solver with respect to grid resolution was tested using grids with $(N_x,N_y)=(50,64)$, $(100,64)$, and $(64,96)$. The $(100,64)$ grid doubles the radial resolution, while the $(64,96)$ grid increases the poloidal resolution from 8 to 12 planes in each divertor region. As shown in Figure~\ref{fig:grid_resolution}, the outer-midplane $N_e$ and $T_e$ profiles and the outer-target $N_e$ profile are relatively insensitive to the tested grid resolutions, whereas the total heat flux incident on the outer target shows a stronger sensitivity to the poloidal resolution.

Some differences are expected because the grids do not have identical radial boundary locations. For technical reasons, the prescribed poloidal-flux values used during mesh generation include the boundary cells, which can introduce small mismatches between grids \citep{dudson2024hermes}. This issue will be addressed in future work. Nevertheless, the solver remained robust with the STRUMPACK-based ILU preconditioner and reached steady state on the higher-resolution grids in less than 30 minutes using 16 CPUs, starting from the same initial conditions as those used in Section~\ref{sec:SolverOptimization}. The results presented in Section~\ref{sec:Results} were obtained using the $(N_x,N_y)=(64,96)$ grid.


\section{Application} \label{sec:Results}

In this section, the results are presented and discussed. The improvements to the PETSc SNES-based steady-state solver described in Section~\ref{sec:SolverOptimization} enabled the characterization of divertor detachment in DIII-D. The detached divertor conditions are first characterized, followed by an analysis of the steady-state particle and energy balances. The effects of varying the upstream separatrix electron density are then investigated, followed by an assessment of the sensitivity of the results to different parallel heat-flux limiter models.


\subsection{Discharge characterization}
\label{sec:discharge_characterization}

The detached regime is characterized by strong plasma cooling in the divertor region due to radiation from impurities, which in this case is dominated by N$_2$. While detailed impurity transport can be important for accurately reproducing experimental conditions, simplified approaches, such as the fixed-fraction radiation model in which the impurity level is prescribed as a fraction of the electron density, have been shown to capture the essential physics of detachment \citep{rognlien2017comparison}. Furthermore, the spatial distribution of intrinsic impurities, such as carbon, may have a limited impact on the overall detachment dynamics \citep{zhao20242d}.


The radiative power loss in the fixed-fraction radiation model is given by $P_{\mathrm{rad}} = - f_{\mathrm{imp}} L_{\mathrm{imp}} n_e^2$, where $f_{\mathrm{imp}}$ is the impurity fraction, $L_{\mathrm{imp}}$ is the radiation loss function, and $n_e$ is the electron density. The impurity radiation loss function is set to zero inside the separatrix. This discharge corresponds to a relatively high-$Z_{\mathrm{eff}}$ case, so three nitrogen impurity fractions, 1\%, 3\%, and 6\%, were investigated. Results are therefore presented for these three cases. The nitrogen radiation loss function was obtained from the ADAS database \citep{summers2011atomic}. In addition, a constant carbon fraction of 2\% was assumed, following \cite{rognlien2017comparison}, with the corresponding radiation loss function taken from \cite{hutchinson1994thermal}.

\begin{figure}[!ht]
\centering
\includegraphics[width=0.49\linewidth]{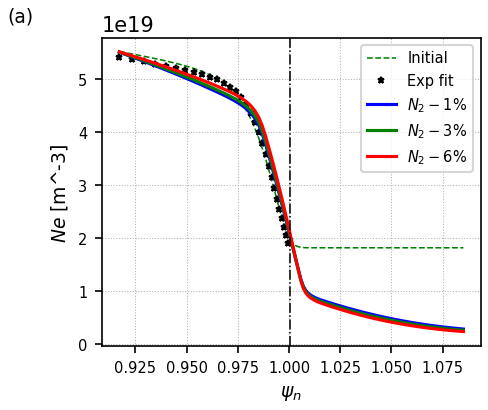}
\includegraphics[width=0.49\linewidth]{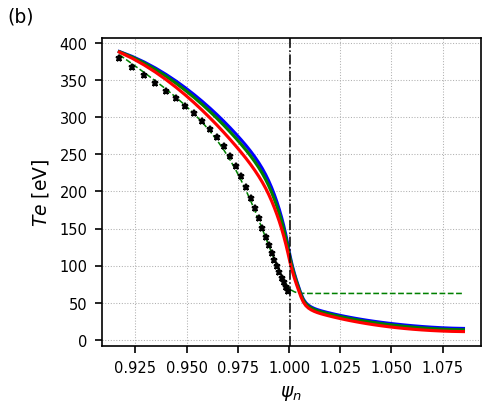}
\includegraphics[width=0.49\linewidth]{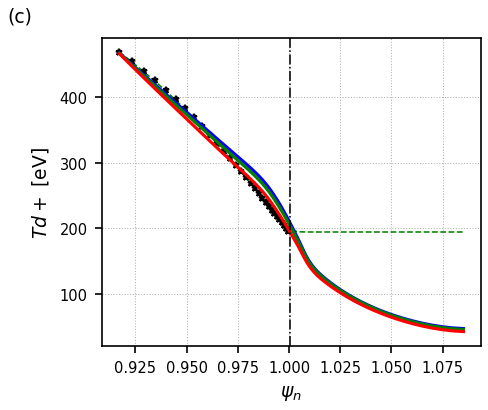}
\includegraphics[width=0.49\linewidth]{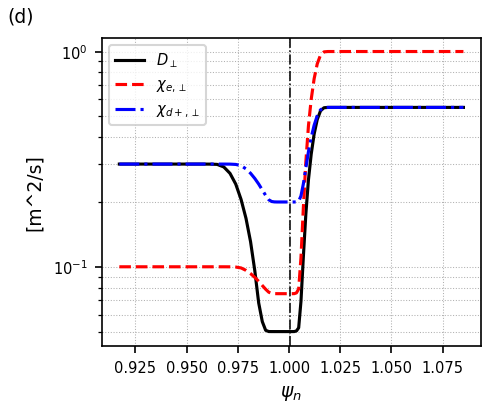}
\caption{Outer-midplane profiles: (\textit{a}) electron density, $N_e$; (\textit{b}) electron temperature, $T_e$; (\textit{c}) deuterium ion temperature, $T_{d+}$; and (\textit{d}) anomalous cross-field transport coefficients ($D_\perp$, $\chi_{e,\perp}$, $\chi_{d+,\perp}$). Profiles are shown for three impurity fractions, 1\%, 3\%, and 6\% N$_2$. Fits to the experimental data are shown as black markers, while the initial conditions used in the simulations are indicated by green dashed lines. The vertical dash-dotted line indicates the location of the separatrix.}
\label{fig:discharge_profiles}
\end{figure}

Figure~\ref{fig:discharge_profiles} presents results from the two-dimensional Hermes-3 simulations. Panels (a)--(c) show the outer-midplane profiles of the electron density, electron temperature, and deuterium ion temperature. Fits to the experimental profiles from the kinetic-EFIT reconstruction of the DIII-D discharge are overlaid as black markers, while the initial profiles used in the simulations (see Section~\ref{sec:ComputationalSetUp}) are shown as green dashed lines. The model assumes quasineutrality, so the deuterium ion and electron densities are equal. This approximation may lead to discrepancies with experimental total-pressure measurements and will be addressed in future multi-species simulations.

The anomalous cross-field transport coefficients, representing turbulent transport, are tuned to reproduce the experimentally measured density and temperature profiles inside the separatrix at the outer midplane. The values providing the best agreement are shown in Figure~\ref{fig:discharge_profiles}(d). The particle diffusivity $D_\perp$ is parameterized in three regions—core ($D_{\mathrm{core}}$), separatrix ($D_{\mathrm{sep}}$), and scrape-off layer ($D_{\mathrm{SOL}}$)—with smooth transitions enforced via hyperbolic tangent functions. The same parameterization is applied to the electron and ion heat diffusivities, $\chi_{e,\perp}$ and $\chi_{d+,\perp}$. These transport coefficients are assumed to be poloidally uniform, such that the same radial profiles are applied at all poloidal locations.

As illustrated in Figure~\ref{fig:discharge_profiles}(d), matching the experimental radial profiles requires reducing $D_\perp$ by approximately an order of magnitude within the edge transport barrier, characteristic of H-mode plasmas. In contrast, the reduction in heat diffusivities is more modest, consistent with \cite{rozhansky2009new}. This three-region parameterization yields good agreement with the measured profiles of $n_e$ and $T_{d+}$ inside the separatrix, while $T_e$ is somewhat overestimated in the region $\psi_n = 0.97 - 1$. Alternative approaches for tuning transport coefficients, such as those proposed in \cite{li2018calculation} and the iterative method of \cite{canik2011measurements}, were also explored; however, for the present discharge, characterized by steep density gradients, these methods did not produce converged solutions. Future work could explore the use of statistical fitting methods~\citep{fu2026uedge}.

In the SOL, where experimental data are typically limited, the transport coefficients are treated as model parameters. Recent parametric studies by \cite{xu2025fluctuation} show that increasing $D_{\mathrm{SOL}}$ and the corresponding heat diffusivities $\chi_{\perp}$ enhances radial transport, broadens the heat-flux width at the target plate, and may facilitate divertor plasma detachment.

Figure~\ref{fig:discharge_profiles} shows the outer-midplane profiles for three nitrogen impurity fractions, $f_{\mathrm{imp}}=1\%$, 3\%, and 6\%, denoted in the figure as 1\%, 3\%, and 6\% N$_2$, respectively. Only modest differences are observed among the three cases. This behavior is expected because the radiation loss functions for both nitrogen and carbon peak at electron temperatures of approximately 12\,eV. In contrast, electron temperatures near the separatrix exceed $\sim 80$\,eV, where impurity radiation is much less effective. From a modeling perspective, the fixed-fraction radiation model acts as a sink term in the electron pressure equation. Upstream, however, parallel electron heat conduction dominates the energy balance. Consequently, the upstream temperature profiles are determined primarily by parallel transport rather than by local radiative losses as will be discussed in Section~\ref{sec:balancing_terms}.

\begin{figure}[!ht]
\centering
\includegraphics[width=0.49\linewidth]{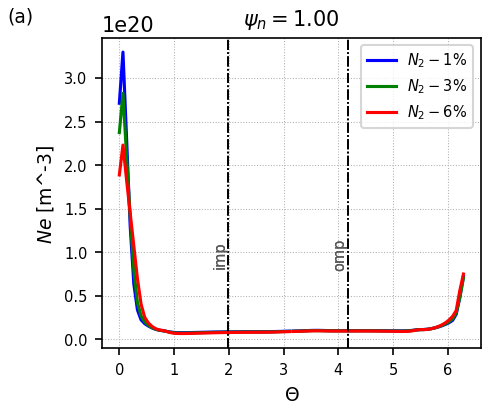}
\includegraphics[width=0.49\linewidth]{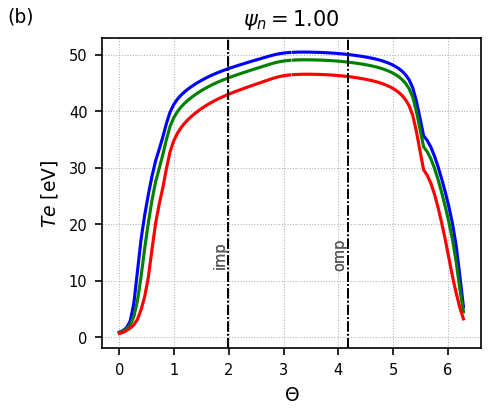}
\caption{Poloidal profiles of the electron density, $N_e$, and electron temperature, $T_e$, near the separatrix, $\psi_n=1$. Profiles are shown for three impurity fractions, 1\%, 3\%, and 6\% N$_2$. The vertical dash-dotted lines indicate the locations of the inner and outer midplanes.}
\label{fig:discharge_profiles_pol}
\end{figure}

\begin{figure}[!ht]
\centering
\includegraphics[width=0.49\linewidth]{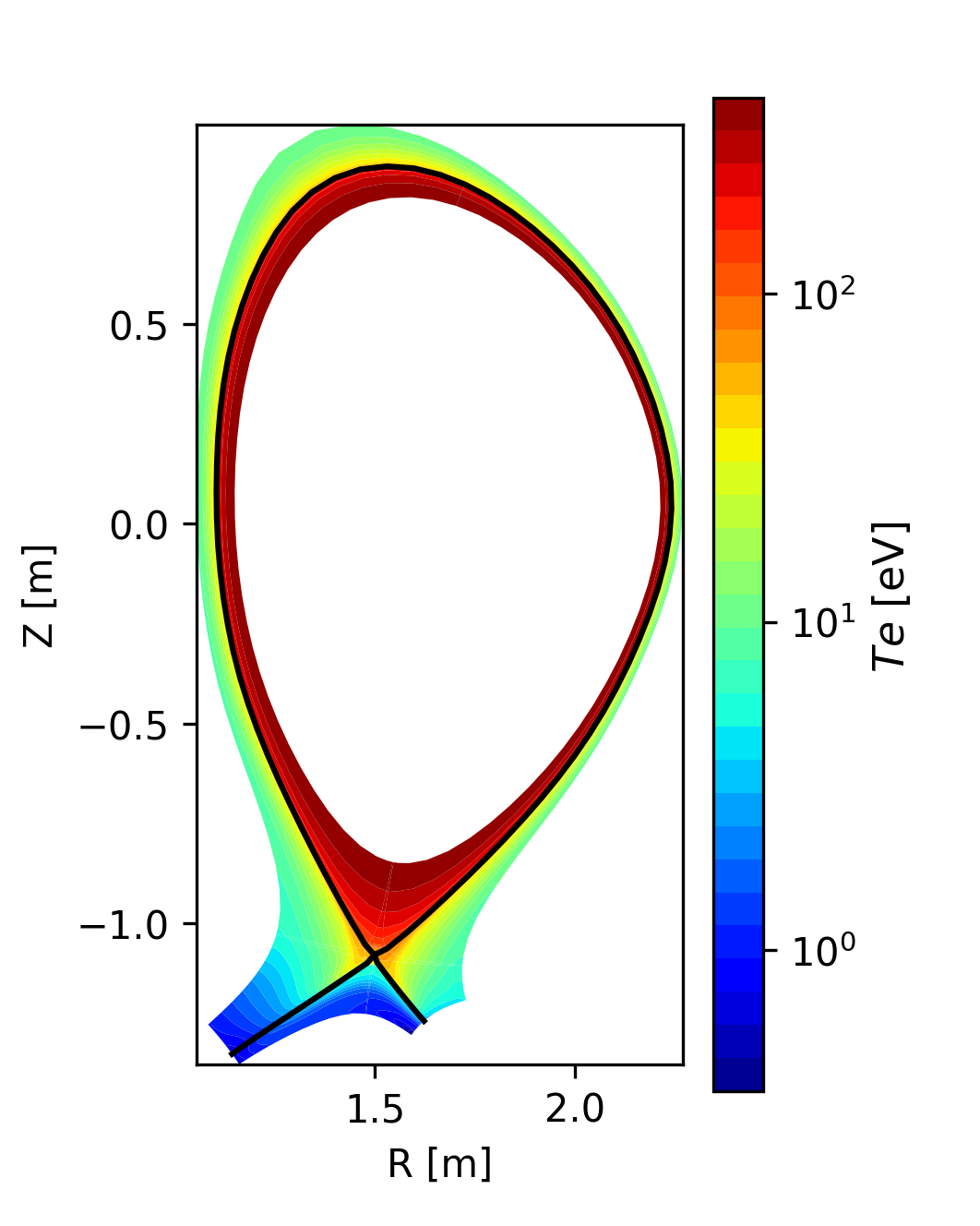}
\includegraphics[width=0.49\linewidth]{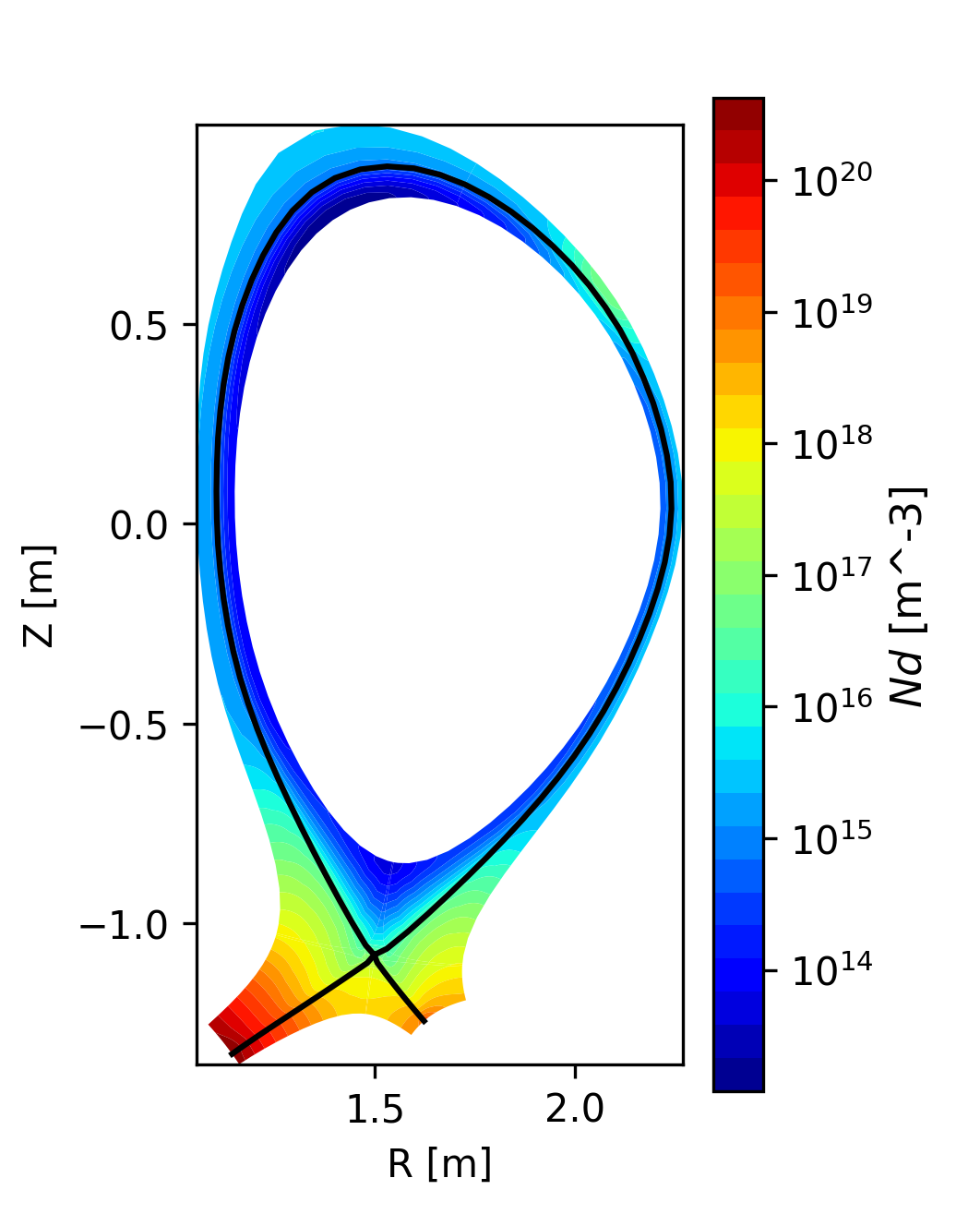}
\caption{Two-dimensional contour plots of the steady-state electron temperature, $T_e$, and neutral density, $N_d$. Results are shown for the 6\% N$_2$ case. The solid black curve denotes the location of the last closed flux surface (separatrix).}
\label{fig:discharge_profiles_2d}
\end{figure}

The poloidal distribution of electron density and temperature near the separatrix for three impurity fractions (1\%, 3\%, and 6\% N$_2$) is shown in Figure~\ref{fig:discharge_profiles_pol}(a) and (b), respectively. The profiles are plotted as a function of the poloidal angle $\Theta$, ranging from 0 to $2\pi$, where $\Theta=0$ corresponds to the inner target and $\Theta=2\pi$ to the outer target for this single-null configuration. The electron temperature, $T_e$, decreases from the midplane toward the divertor due to energy losses in the divertor legs, while the electron density, $N_e$, increases toward the targets as a result of recycling of incident ions into neutrals and their subsequent ionization. 
Consequently, the electron pressure, and similarly the ion pressure, remains relatively constant upstream and decreases in the divertor as a result of energy and momentum losses. As discussed in more detail in Section~\ref{sec:balancing_terms}, electrons in the outer divertor lose energy primarily through ionization and impurity radiation, while ions lose energy through collisions with electrons.

The inner divertor is well detached in all cases, with $T_e$ reaching values of approximately 1~eV. In contrast, the outer divertor remains at higher temperatures, with $T_e$ decreasing to approximately 5~eV only near the target. This behavior is further illustrated in the two-dimensional contour plots shown in Figure~\ref{fig:discharge_profiles_2d}, where $T_e$ and the neutral density, $N_d$, are presented for the 6\% N$_2$ case. In particular, the region with $T_e \sim 1$\,eV is broader in the inner divertor than in the outer divertor and is accompanied by higher neutral densities. As a result, the peak heat flux to the inner target is approximately $1.2~\mathrm{MW/m^2}$, corresponding to a reduction by nearly one order of magnitude relative to the outer-target value of approximately $11~\mathrm{MW/m^2}$.

\begin{figure}[!ht]
\centering
\includegraphics[width=0.49\linewidth]{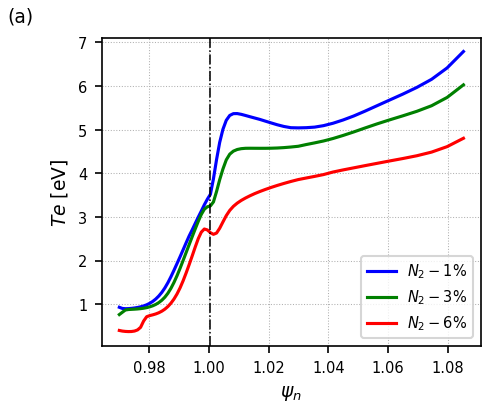}
\includegraphics[width=0.49\linewidth]{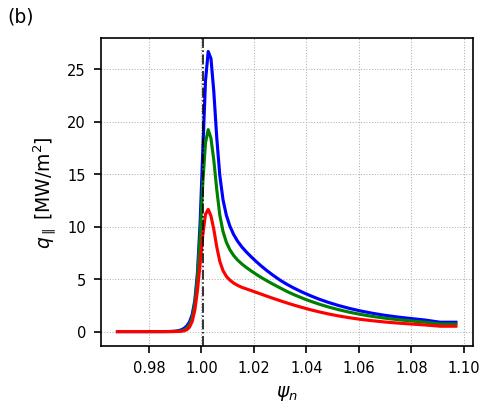}
\caption{Profiles at the outer target plate: (\textit{a}) electron temperature, $T_e$;   (\textit{b}) total parallel heat flux, $q_\parallel$. Profiles are shown for three impurity fractions, 1\%, 3\%, and 6\% nitrogen. The vertical dash-dotted line indicates the location of the separatrix.}
\label{fig:discharge_profiles_target}
\end{figure}

Figure~\ref{fig:discharge_profiles_target} presents the electron temperature and total parallel heat flux at the outer target plate for three impurity fractions, 1\%, 3\%, and 6\% N$_2$. The total parallel heat flux is computed as the sum of the electron and deuterium ion contributions, given by $q_{\parallel,e,d+} = \gamma_{e,d+} n_e T_{e,d+} c_s$. For the 1\% nitrogen case, the electron temperature at the outer strike point reaches approximately 5.3\,eV, with a corresponding peak heat flux of $\sim 27~\mathrm{MW/m^2}$. In contrast, the 6\% nitrogen case, corresponding to stronger divertor radiation, results in target electron temperatures below 5\,eV and a reduced peak heat flux of approximately $11~\mathrm{MW/m^2}$. 

It should be noted that a divertor electron temperature of $\sim 5$\,eV is commonly used as an indicator of detachment \citep{militello2022boundary}, while \cite{stangeby2018basic} identifies $T_{e,\mathrm{target}} \approx 10$\,eV as the approximate transition between the high-recycling and detached divertor regimes. In addition, reactor material constraints generally require steady-state target heat fluxes below $10~\mathrm{MW/m^2}$ \citep{stangeby2018basic}, a condition almost achieved here for the 6\% nitrogen case. Detachment is also frequently associated with the rollover of the divertor ion flux \citep{leonard2018plasma}. An upstream separatrix density scan was performed to examine the response of the model and is discussed in Section~\ref{sec:scan_ne}.

\subsection{Steady-state particle and energy balances} \label{sec:balancing_terms}

To further characterize the converged plasma state, the individual contributions to the deuterium ion density, ion pressure, and electron pressure equations are evaluated separately for the case with a 6\% N$_2$ impurity fraction presented in Section~\ref{sec:discharge_characterization}. At steady state, the time derivatives approach zero, and the remaining source, sink, and transport terms must therefore balance. In addition to providing a numerical check of the steady-state solution, this analysis identifies the dominant physical mechanisms in different regions of the plasma. 

\begin{figure}[!ht]
\centering
\includegraphics[width=0.90\linewidth]{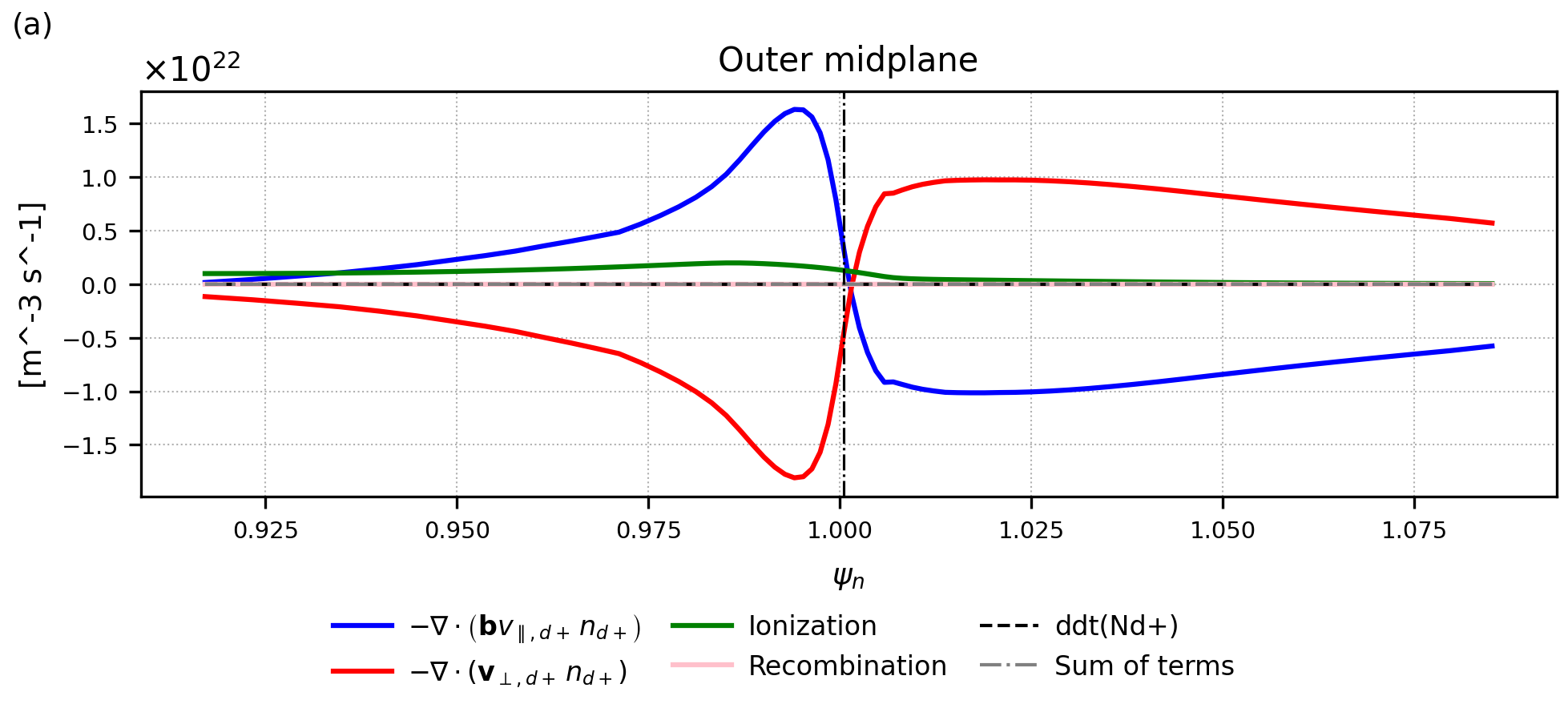}
\includegraphics[width=0.90\linewidth]{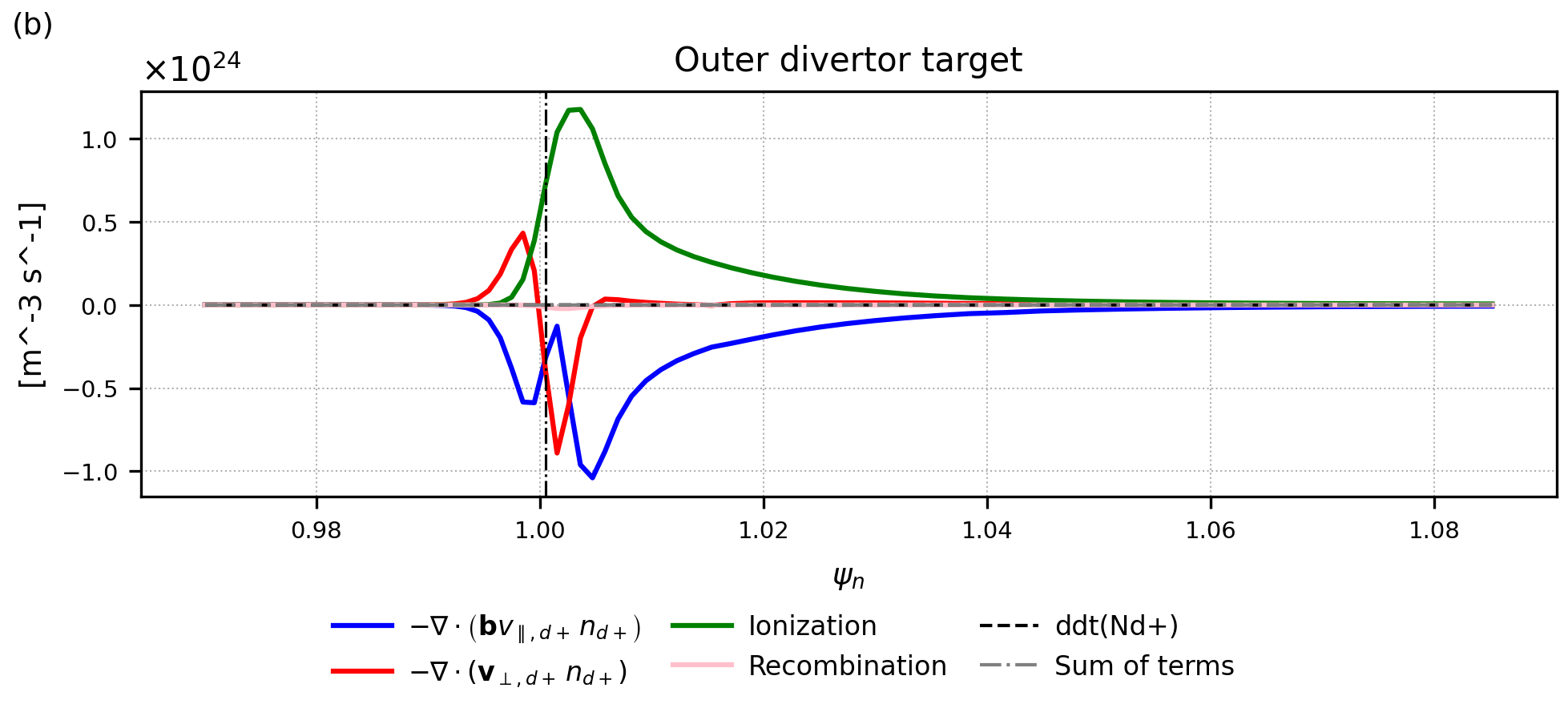}
\includegraphics[width=0.90\linewidth]{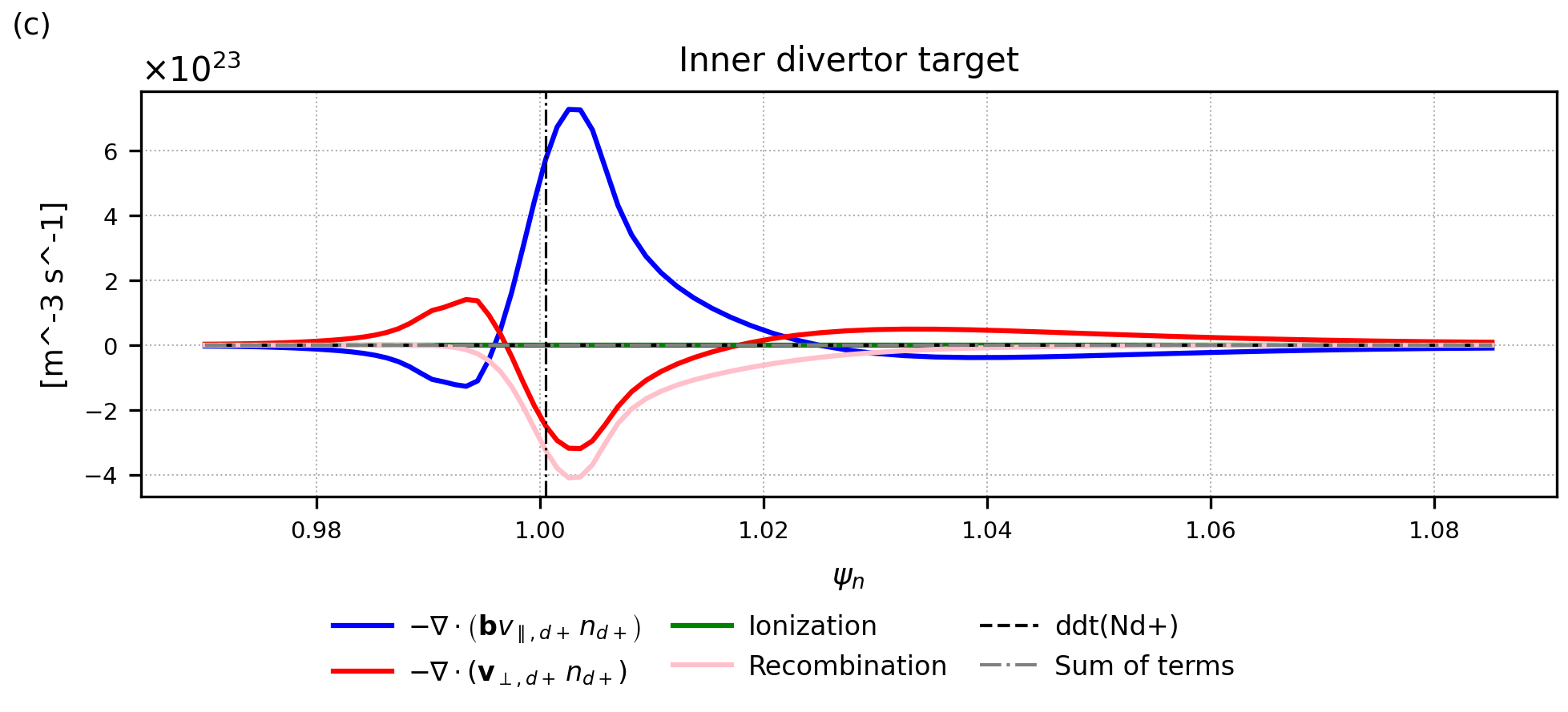}
\caption{Steady-state balance of terms in Equation~\eqref{eq:ion_number_density} for the deuterium ion density equation at (a) the outer midplane, (b) the outer divertor target, and (c) the inner divertor target. The individual transport and atomic-physics contributions are shown together with the time derivative, $\partial n_{d+}/\partial t$ (black dashed), and the sum of the individual terms (gray dash-dotted). Profiles are shown for the case with a 6\% N$_2$ impurity fraction. The vertical line indicates the separatrix.}
\label{fig:balance_density}
\end{figure}

Figure~\ref{fig:balance_density} shows the balance of terms in Equation~\eqref{eq:ion_number_density} for the deuterium ion density at (a) the outer midplane, (b) the outer divertor target, and (c) the inner divertor target. The black dashed line shows the steady-state time derivative, $\partial n_{d+}/\partial t$, while the gray dash-dotted line shows the sum of the individual terms. Both remain close to zero, demonstrating that the different contributions to the density equation balance at steady state.

At the outer midplane, the dominant balance changes across the separatrix. Inside the separatrix, parallel advection acts as a positive contribution to $n_{d+}$ and is mainly balanced by perpendicular anomalous transport. Ionization also provides a positive source due to neutrals penetrating into the hotter plasma, although its contribution is comparatively small. In the SOL, the roles of the two dominant transport terms are reversed: perpendicular anomalous transport supplies particles radially into the SOL, while parallel advection removes them along open field lines toward the divertor targets.

At the outer target, ionization is the dominant particle source and is largely balanced by parallel advection toward the target plate. Perpendicular anomalous transport is mainly important near the strike point, where the density gradients are largest, while recombination remains relatively weak in this partially detached region. In contrast, at the inner divertor target, where the plasma is fully detached and $T_{e,\mathrm{target}}$ falls below 1~eV, recombination becomes the dominant particle loss mechanism. This loss is primarily balanced by parallel advection, while ionization plays a less significant role.

\begin{figure}[!ht]
\centering
\includegraphics[width=0.90\linewidth]{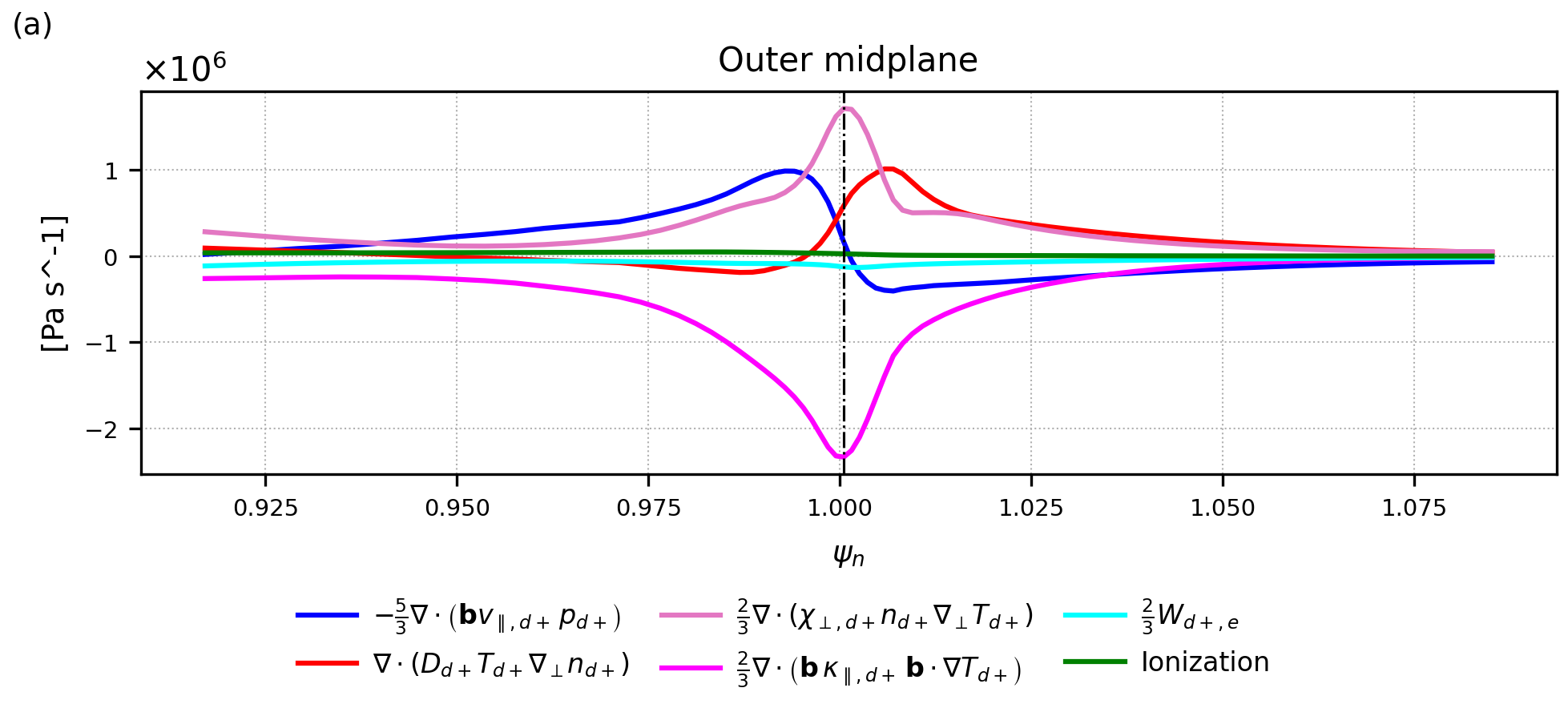}
\includegraphics[width=0.90\linewidth]{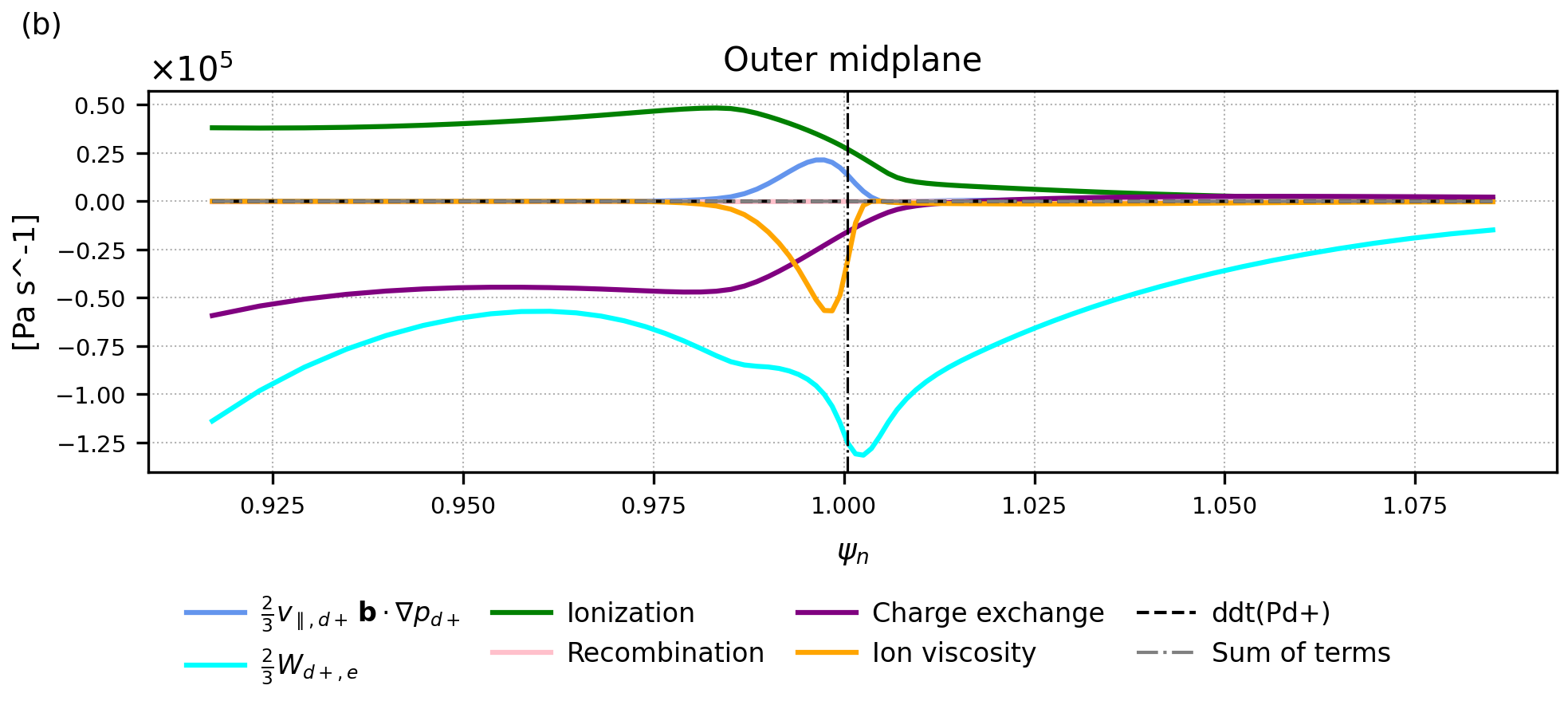}
\caption{Steady-state balance of terms in Equation~\eqref{eq:ion_pressure} for the deuterium ion pressure equation at the outer midplane. Panel (a) shows the dominant contributions, while panel (b) shows the smaller contributions on an expanded scale. The time derivative, $\partial p_{d+}/\partial t$, is shown with a black dashed line and the sum of all terms with a gray dash-dotted line. Profiles are shown for the case with a 6\% nitrogen impurity fraction. The vertical line indicates the separatrix.}
\label{fig:balance_pressure_omp}
\end{figure}


Figure~\ref{fig:balance_pressure_omp} shows the steady-state balance of terms in Equation~\eqref{eq:ion_pressure} for the deuterium ion pressure at the outer midplane. 
For this balance analysis, Equation~\eqref{eq:ion_pressure} is multiplied by $2/3$ so that the plotted terms sum directly to $\partial p_{d+}/\partial t$.
Panel (a) shows the dominant contributions, while panel (b) shows the smaller terms on an expanded scale. Parallel heat conduction, $\frac{2}{3}\nabla \cdot \left( \mathbf{b},\kappa_{\parallel,d+},\mathbf{b}\cdot\nabla T_{d+}\right)$, is the dominant loss mechanism and contributes negatively both inside the separatrix and in the SOL. In the confined region, this loss is primarily balanced by parallel convective transport, $-\frac{5}{3}\nabla\cdot\left(\mathbf{b}v_{\parallel,d+}p_{d+}\right)$, and perpendicular anomalous thermal transport, $\frac{2}{3}\nabla\cdot\left(\chi_{\perp,d+}n_{d+}\nabla_{\perp}T_{d+}\right)$. The anomalous particle-diffusion contribution, $\nabla\cdot\left(D_{\perp,d+}T_{d+}\nabla_{\perp}n_{d+}\right)$, behaves similarly to the corresponding term in the $n_{d+}$ equation, acting as a sink inside the separatrix and as a source in the SOL. By contrast, the anomalous thermal transport term remains positive across the radial domain. The parallel pressure-advection term, $\frac{2}{3}v_{\parallel,d+}\nabla_{\parallel}p_{d+}$, makes only a small contribution upstream.

The smaller terms are shown in Figure~\ref{fig:balance_pressure_omp}(b). Ionization provides a positive contribution to the ion pressure, while electron-ion collisional energy exchange, $W_{d+,e}$, is negative because $T_{d+}>T_e$, resulting in energy transfer from ions to electrons. Charge exchange is also negative, indicating energy transfer from ions to neutrals, whereas recombination is negligible at the outer midplane. The ionization, recombination, and charge-exchange terms include both temperature-dependent contributions, e.g. $\frac{3}{2} R_{cx} e T_d$, and relative-velocity contributions, e.g. $\frac{1}{2} m_d R_{cx} (v_{\parallel,d}-v_{\parallel,d+})^2$; the latter are much smaller in the present simulations. Parallel ion viscosity contributes only weakly, mainly near the separatrix.

\begin{figure}[!ht]
\centering
\includegraphics[width=0.90\linewidth]{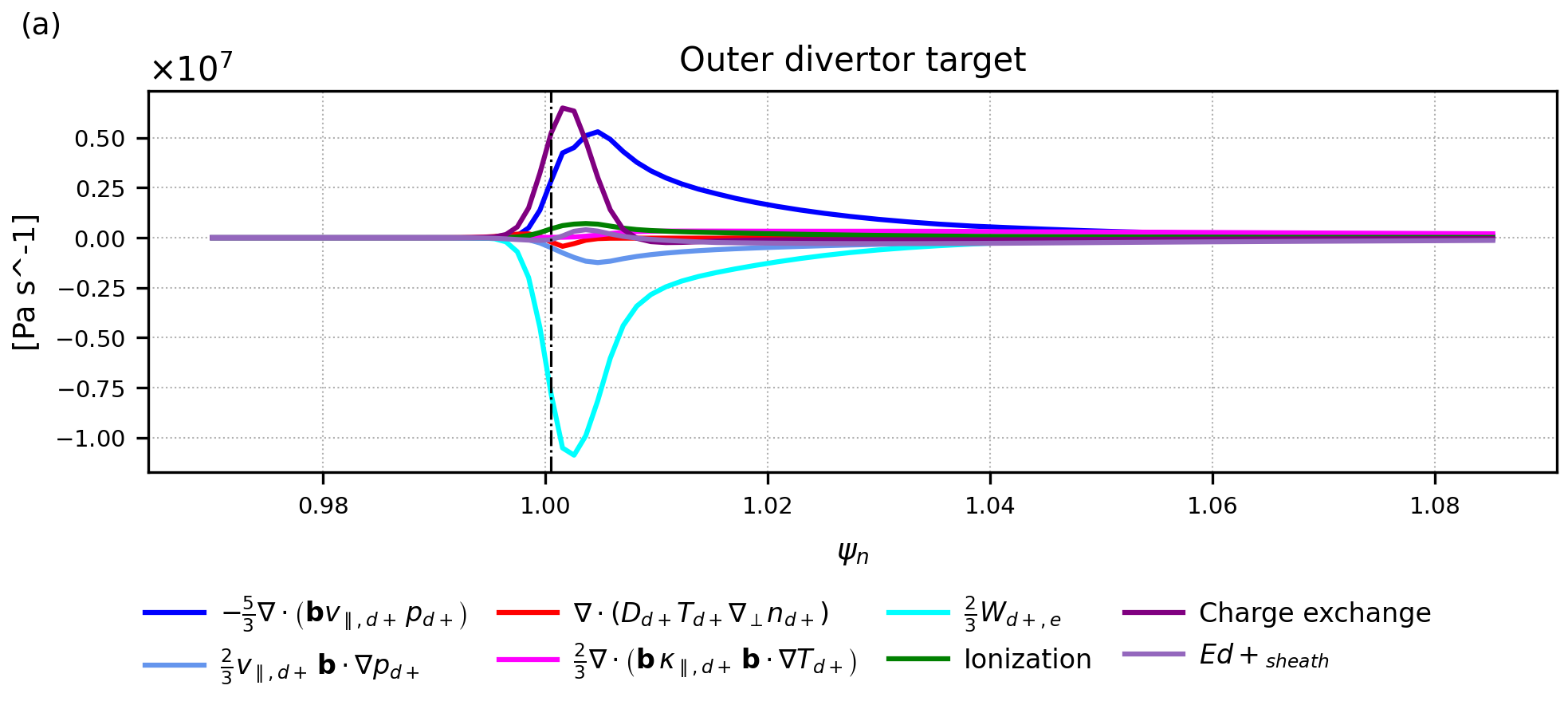}
\includegraphics[width=0.90\linewidth]{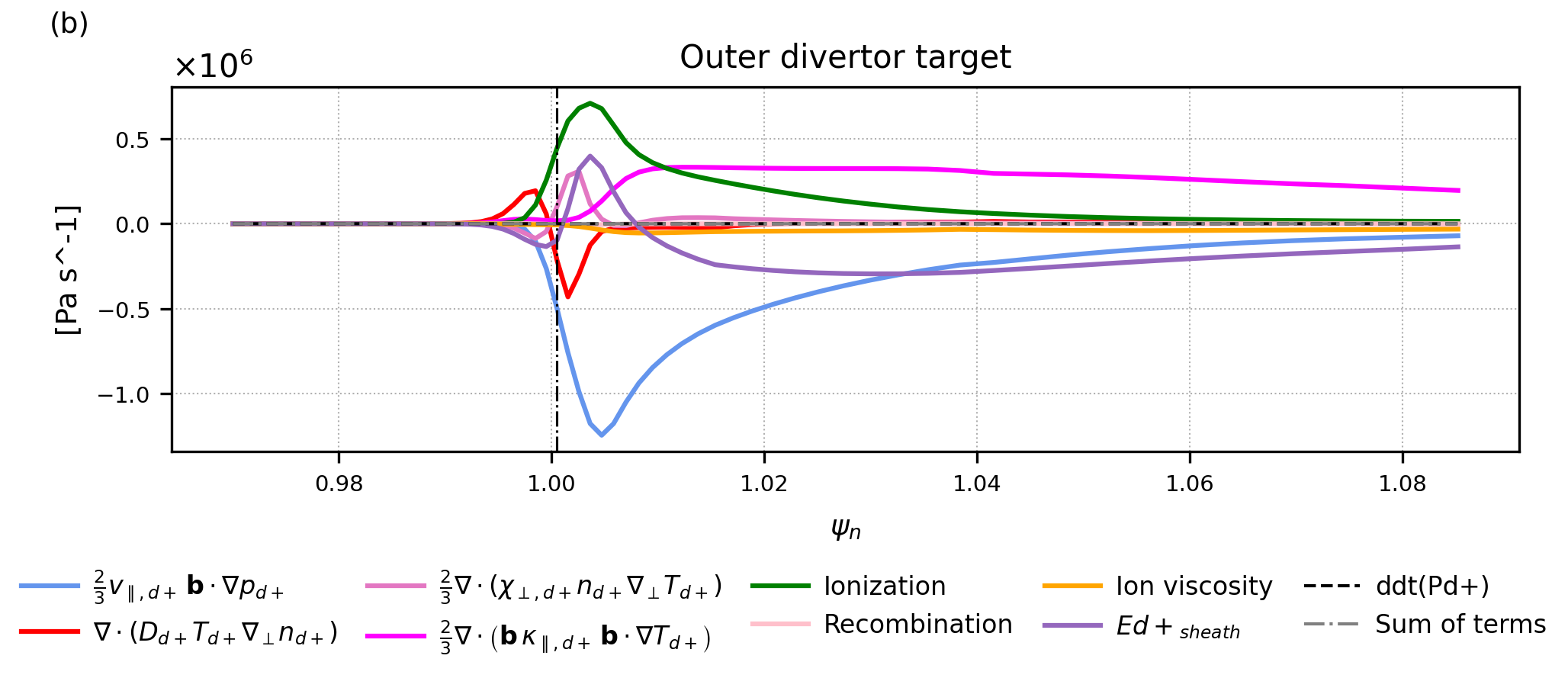}
\caption{Steady-state balance of terms in Equation~\eqref{eq:ion_pressure} for the deuterium ion pressure equation at the outer divertor target. Panel (a) shows the dominant contributions, while panel (b) shows the smaller contributions on an expanded scale. The time derivative, $\partial p_{d+}/\partial t$, is shown with a black dashed line and the sum of all terms with a gray dash-dotted line. Profiles are shown for the case with a 6\% N$_2$ impurity fraction. The vertical line indicates the separatrix.}
\label{fig:balance_pressure_OT}
\end{figure}

Figure~\ref{fig:balance_pressure_OT} shows the steady-state deuterium ion pressure balance at the outer divertor target, with the dominant terms shown in panel (a) and the smaller contributions in panel (b). The balance is qualitatively different from that at the outer midplane. The dominant loss mechanism is electron-ion collisional energy exchange, $W_{d+,e}$, which is negative because $T_{d+}>T_e$, resulting in energy transfer from ions to electrons. This loss is mainly balanced by parallel convective transport and charge exchange. In contrast to the upstream region, charge exchange contributes positively because the neutral temperature, $T_d$, is slightly higher than $T_{d+}$. Parallel heat conduction is less important in the cooler divertor plasma, although it becomes dominant in the far SOL, $\psi_n>1.05$, where it contributes positively. The parallel pressure-advection term is negative and more significant than upstream due to the larger parallel pressure gradients in the divertor. The sheath contribution can take either sign because, in Hermes-3, it represents a correction to the energy flux already carried through the sheath by the pressure and momentum equations, rather than the total sheath energy loss itself \citep{hermes3_docs}. This correction is applied such that the ion energy flux to the sheath is given by $q_{\parallel,i} = \gamma_{d+} n_{d+} T_{d+} c_s$.

\begin{figure}[!ht]
\centering
\includegraphics[width=0.90\linewidth]{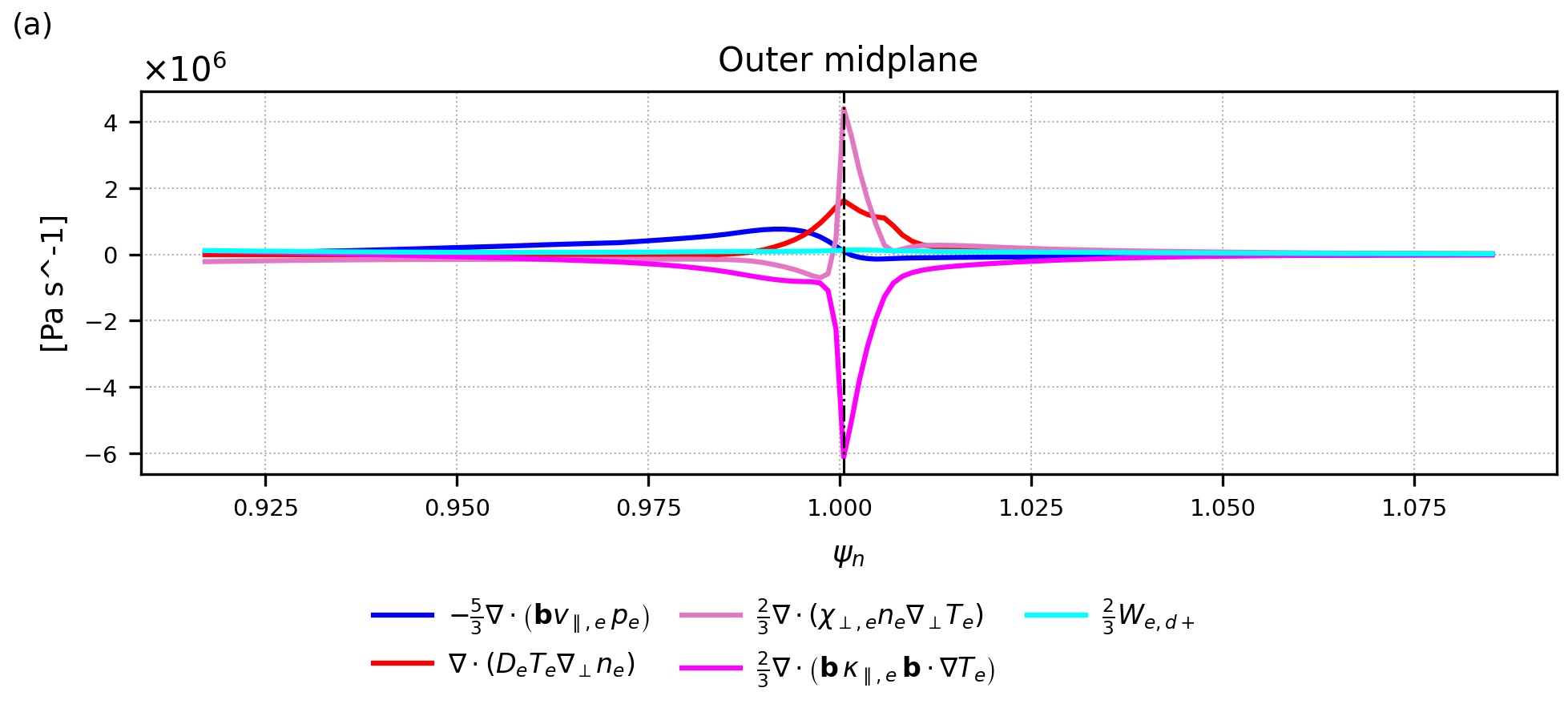}
\includegraphics[width=0.90\linewidth]{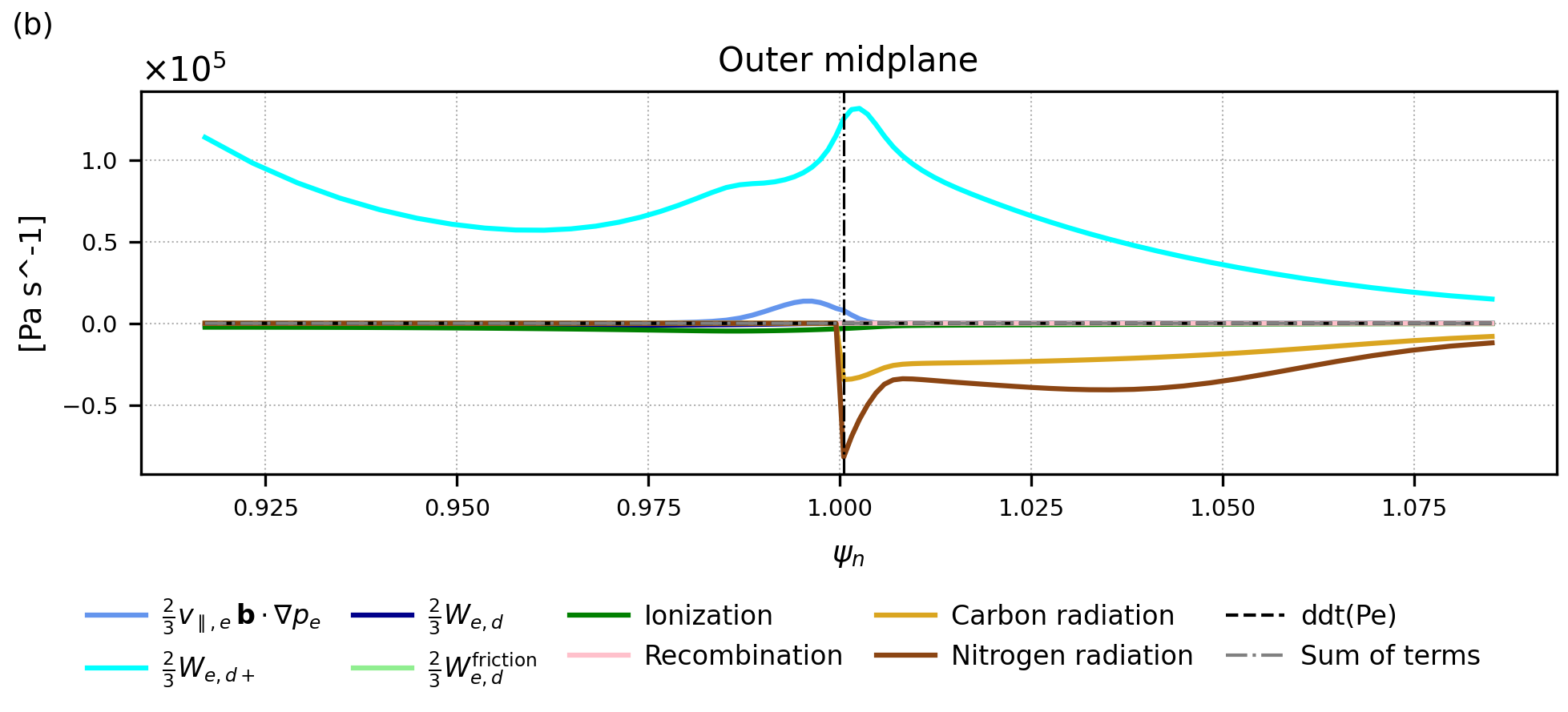}
\caption{Steady-state balance of terms in Equation~\eqref{eq:electron_pressure} for the electron pressure equation at the outer midplane. Panel (a) shows the dominant contributions, while panel (b) shows the smaller contributions on an expanded scale. The time derivative, $\partial p_e/\partial t$, is shown with a black dashed line and the sum of all terms with a gray dash-dotted line. Profiles are shown for the case with a 6\% nitrogen impurity fraction. The vertical line indicates the separatrix.}
\label{fig:balance_e_pressure_omp}
\end{figure}

Figure~\ref{fig:balance_e_pressure_omp} shows the steady-state balance of terms in Equation~\eqref{eq:electron_pressure} for the electron pressure at the outer midplane. Equation~\eqref{eq:electron_pressure} is multiplied by $2/3$ so that the plotted terms sum directly to $\partial p_e/\partial t$. Panel (a) shows the dominant contributions, while panel (b) shows the smaller terms on an expanded scale. The balance is qualitatively similar to that of the deuterium ions. The dominant heat-loss mechanism is parallel heat conduction, $\frac{2}{3}\nabla \cdot \left(\mathbf{b},\kappa_{\parallel,e},\mathbf{b}\cdot\nabla T_e\right)$. This loss is primarily balanced by perpendicular anomalous thermal transport, $\frac{2}{3}\nabla\cdot\left(\chi_{\perp,e}n_e\nabla_{\perp}T_e\right)$, and the anomalous particle-diffusion contribution, $\nabla\cdot\left(D_{\perp,e}T_e\nabla_{\perp}n_e\right)$. In the confined region, parallel convective transport, $-\frac{5}{3}\nabla\cdot\left(\mathbf{b}v_{\parallel,e}p_e\right)$, also provides an important positive contribution to the balance.

The smaller terms are shown in Figure~\ref{fig:balance_e_pressure_omp}(b). Electron-ion collisional energy exchange, $W_{e,d+}$, is equal in magnitude and opposite in sign to the corresponding ion term, $W_{d+,e}$, discussed above. Since energy is transferred from ions to electrons in this region, $W_{e,d+}$ is positive. Energy losses due to carbon and nitrogen impurity radiation, calculated using the fixed-fraction radiation model, are also shown in Figure~\ref{fig:balance_e_pressure_omp}(b). These losses are relatively weak upstream, consistent with the discussion in Section~\ref{sec:discharge_characterization}, and are set to zero inside the separatrix in the present model. Energy exchange associated with electron-neutral collisions, $W_{e,d}$ and $W^{\mathrm{friction}}_{e,d}$, is included for completeness but has a negligible contribution to the upstream electron pressure balance. As in the ion pressure balance, both the time derivative and the sum of all terms remain close to zero.

\begin{figure}[!ht]
\centering
\includegraphics[width=0.90\linewidth]{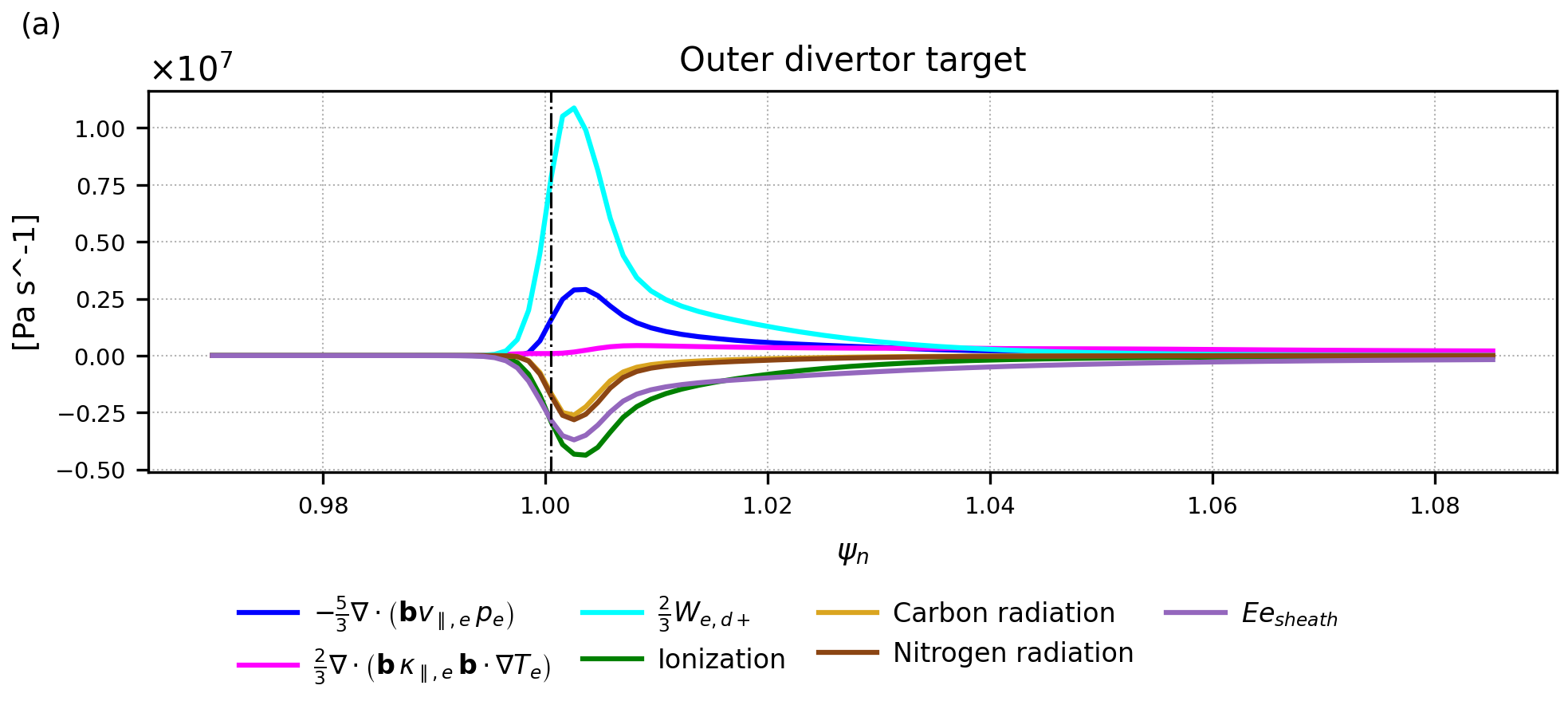}
\includegraphics[width=0.90\linewidth]{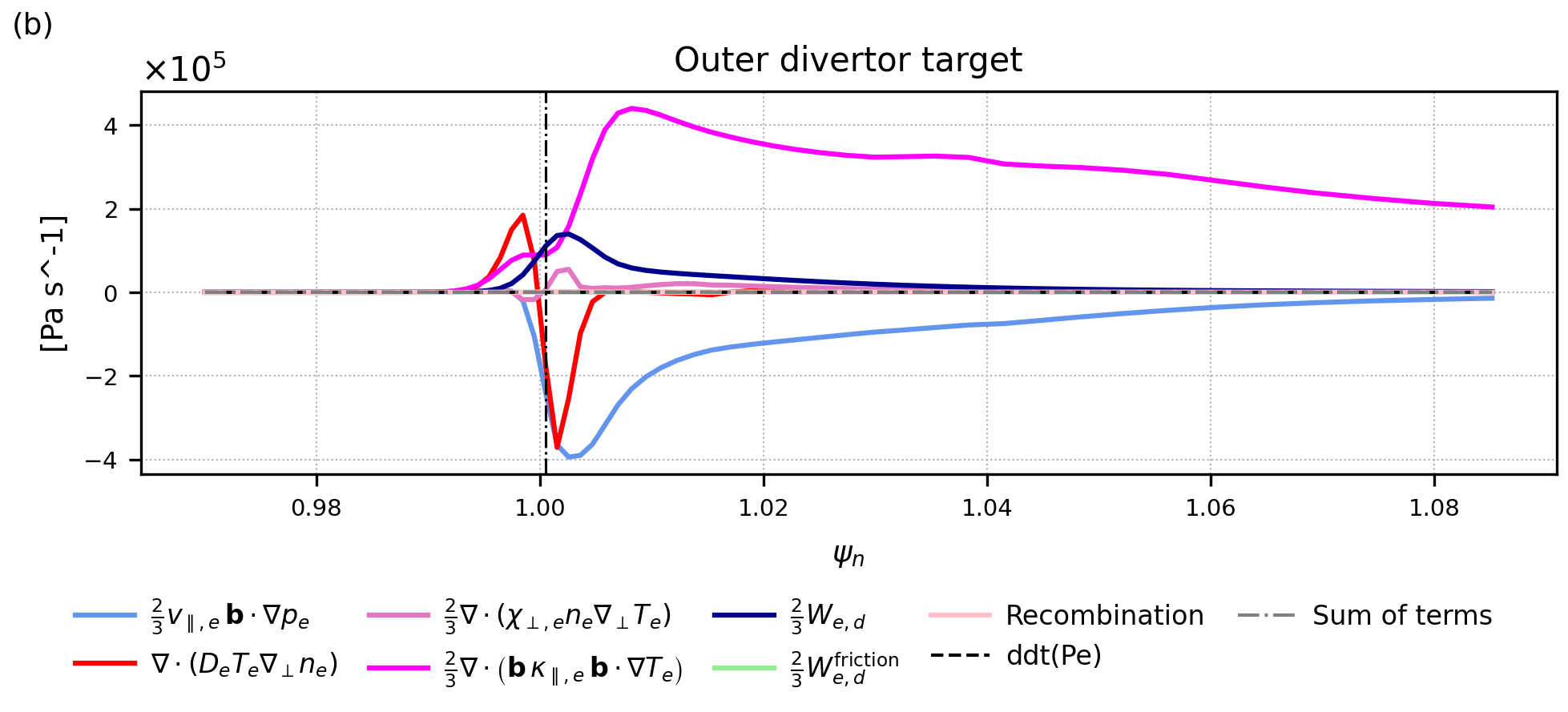}
\caption{Steady-state balance of terms in Equation~\eqref{eq:electron_pressure} for the electron pressure equation at the outer divertor target. Panel (a) shows the dominant contributions, while panel (b) shows the smaller contributions on an expanded scale. The time derivative, $\partial p_e/\partial t$, is shown with a black dashed line and the sum of all terms with a gray dash-dotted line. Profiles are shown for the case with a 6\% N$_2$ impurity fraction. The vertical line indicates the separatrix.}
\label{fig:balance_e_pressure_OT}
\end{figure}

Figure~\ref{fig:balance_e_pressure_OT} shows the steady-state balance of terms in the electron pressure equation at the outer divertor target, with the dominant terms shown in panel (a) and the smaller contributions in panel (b). The dominant positive contributions are electron-ion collisional energy exchange, $W_{e,d+}$, corresponding to energy transfer from ions to electrons, and parallel convective transport, $-\frac{5}{3}\nabla\cdot\left(\mathbf{b}v_{\parallel,e}p_e\right)$. The main electron energy losses arise from ionization, sheath losses, and carbon and nitrogen impurity radiation. Here, the ionization and recombination terms represent the associated radiative energy losses, with rates obtained from the AMJUEL database \citep{reiter2000data}. The carbon and nitrogen radiation losses are comparable in magnitude, despite impurity fractions of 2\% and 6\%, respectively.

The smaller terms are shown in Figure~\ref{fig:balance_e_pressure_OT}(b). The anomalous perpendicular transport terms make only a small contribution, mainly because of the weak radial gradients in $T_e$. Parallel heat conduction becomes important only in the far SOL, for $\psi_n>1.05$, while recombination remains negligible. Energy exchange through electron-neutral collisions, $W_{e,d}$, is finite but remains small compared with the dominant contributions.

Taken together, the electron and ion pressure balances provide a consistent picture of the dominant energy-loss pathways in the outer divertor. Electrons lose energy primarily through ionization and impurity radiation, while ions lose energy mainly through collisional energy transfer to electrons.

\subsection{Number density scan} \label{sec:scan_ne}

\begin{figure}[!ht]
\centering
\includegraphics[width=0.49\linewidth]{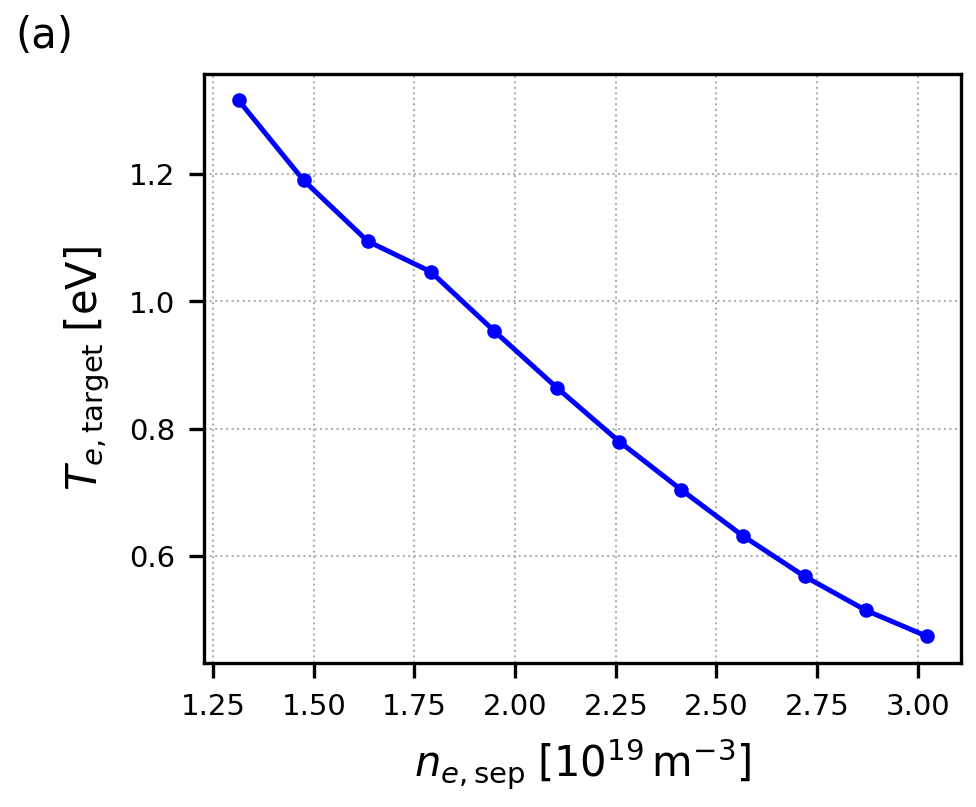}
\includegraphics[width=0.49\linewidth]{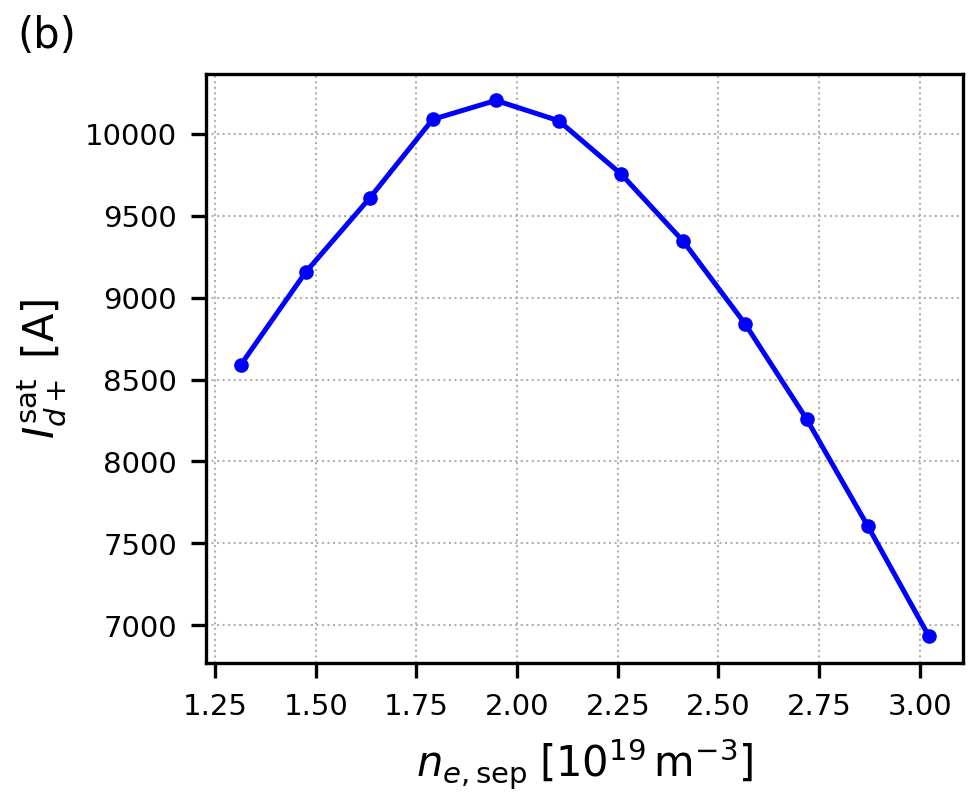}
\caption{Peak electron temperature at the outer divertor target, $T_{e,\mathrm{target}}$, (a) and integrated ion saturation current, $I_{d+}^{\mathrm{sat}}$, (b) as functions of the upstream separatrix electron density, $n_{e,\mathrm{sep}}$. The computational domain spans $\psi_n=0.99$--1.1. Different values of $n_{e,\mathrm{sep}}$ are obtained by varying the core Dirichlet boundary condition for $n_e$.}
\label{fig:Te_target_vs_ne_sep}
\end{figure}

The upstream separatrix electron density is commonly used as a control parameter for divertor detachment. To examine its effect on divertor conditions, a parameter scan is performed by varying the upstream density. Following \cite{rognlien2017comparison,zhang2024uedge}, a shorter computational domain in the radial direction is employed to provide better control of $n_{e,\mathrm{sep}}$ through the imposed core boundary condition for $n_e$. Although the shortened domain may modify local plasma conditions near the separatrix relative to the results presented earlier, it remains useful for investigating overall detachment trends.

Figure~\ref{fig:Te_target_vs_ne_sep}(a) shows the peak electron temperature at the outer divertor target, $T_{e,\mathrm{target}}$, as a function of the upstream separatrix electron density, $n_{e,\mathrm{sep}}$. The computational domain for these simulations spans the range $\psi_n = 0.99$--1.1 with a resolution of $50 \times 64$ in $(x,y)$, while the anomalous transport coefficients are fixed at $D_\perp = 0.2$ and $\chi_{e,\perp} = \chi_{d+,\perp} = 0.55$ throughout the domain, following \cite{rognlien2017comparison}.  An impurity fraction of 4\% N$_2$ was used. As expected, $T_{e,\mathrm{target}}$ decreases with increasing $n_{e,\mathrm{sep}}$, as higher upstream densities enhance divertor volumetric losses, particularly radiation, which scales approximately as $n_e^2$, thereby facilitating detachment.

Detachment is also frequently associated with the rollover of the divertor ion flux, defined as the point at which the target ion flux decreases with increasing upstream separatrix density \citep{leonard2018plasma,militello2022boundary}. Figure~\ref{fig:Te_target_vs_ne_sep}(b) shows the corresponding integrated ion saturation current $I_{d+}^{\mathrm{sat}}$ as a function of $n_{e,\mathrm{sep}}$. The current initially increases with $n_{e,\mathrm{sep}}$, reaches a maximum at approximately $n_{e,\mathrm{sep}}\simeq2\times10^{19}$~m$^{-3}$, and subsequently decreases as the density is increased further. At approximately the same density, the peak outer-target electron temperature falls below 1~eV. At these low temperatures, recombination becomes increasingly important, reducing the ion density reaching the target and contributing to the observed rollover of the ion flux.

A bifurcation-like drop in $T_{e,\mathrm{target}}$ with increasing $n_{e,\mathrm{sep}}$ has been experimentally observed in DIII-D plasmas \citep{mclean2015electron} and reproduced in UEDGE \citep{jaervinen2018b,rognlien2017comparison,zhao20242d} and SOLPS-ITER \citep{du2020solps} simulations. Such a bifurcation-like transition is not observed in Figure~\ref{fig:Te_target_vs_ne_sep}, likely because the present simulations do not include magnetic drifts. Previous studies have identified the effect of $E \times B$ and magnetic drifts as important mechanisms for capturing the detachment cliff dynamics. This topic will therefore be revisited in future Hermes-3 studies including drift effects.

\subsection{Parallel heat-flux limiters} \label{sec:FluxLimiters}

It is useful to examine the sensitivity of detached-divertor simulations to the choice of parallel heat-flux limiter. The collisional Braginskii closure \citep{braginskii1965transport} is not valid in weakly collisional or collisionless regimes and can substantially overestimate the parallel conductive heat flux in the confined region and, under some conditions, near the separatrix \citep{zhu2021drift,zholobenko2024tokamak}. Alternative closures and flux-limiting models have therefore been proposed \citep{hammett1990fluid,snyder1997landau,schurtz2000nonlocal,dimits2013fast,umansky2015modeling}.

In the classical Braginskii, or Spitzer--Härm (SH), closure, the parallel conductive heat-flux density for species $j$ is
\begin{equation}
q_{\parallel,j}^{\mathrm{SH}}
=
-\kappa_{\parallel,j}^{\mathrm{SH}}\nabla_{\parallel}T_j,
\qquad
\kappa_{\parallel,j}^{\mathrm{SH}}
=
\begin{cases}
3.16\,\dfrac{p_e}{m_e\nu_e}, & j=e, \\[6pt]
3.9\,\dfrac{p_{d+}}{m_{d+}\nu_{d+}}, & j=d+,
\end{cases}
\label{eq:spitzer_harm_heat_flux}
\end{equation}
where $\nu_j$ is the collision frequency. A common approach is to limit this flux to a fraction of the free-streaming (FS) value, $q_{\parallel,j}^{\mathrm{FS}}=\alpha_j n_j T_j v_{\mathrm{th},j}$, where $v_{\mathrm{th},j}=\sqrt{T_j/m_j}$ and temperatures are expressed in energy units. Here, $\alpha_e=0.3$ and $\alpha_{d+}=1.0$ are used for electrons and deuterium ions, respectively.

Two heat-flux-limiter formulations are considered in the present work. The default model in Hermes-3 is a local, gradient-dependent limiter of the type used in SOLPS \citep{schneider2006plasma,rozhansky2009new}. The limited heat flux is obtained from a harmonic interpolation between the Spitzer--H\"arm and free-streaming values:
\begin{equation}
q_{\parallel,j}
=
\frac{q_{\parallel,j}^{\mathrm{SH}}}
{1+\left|q_{\parallel,j}^{\mathrm{SH}}\right|/
q_{\parallel,j}^{\mathrm{FS}}}.
\label{eq:local_gradient_flux_limiter}
\end{equation}
Equivalently,
\begin{equation}
q_{\parallel,j}
=
-\kappa_{\parallel,j}^{\mathrm{eff}}\nabla_{\parallel}T_j,
\end{equation}
with
\begin{equation}
\kappa_{\parallel,j}^{\mathrm{eff}}
=
\frac{\kappa_{\parallel,j}^{\mathrm{SH}}}
{1+
\dfrac{
\kappa_{\parallel,j}^{\mathrm{SH}}
\left|\nabla_{\parallel}T_j\right|
}{
\alpha_j n_j T_j v_{\mathrm{th},j}
}}.
\label{eq:local_gradient_effective_conductivity}
\end{equation}
This formulation recovers the Spitzer--H\"arm limit at low conductive heat flux and approaches the free-streaming limit at high heat flux. Since the correction depends explicitly on the local parallel temperature gradient, it is referred to here as the local-gradient flux limiter.

As an alternative to the local-gradient model, a connection-length limiter was implemented in Hermes-3 following the free-streaming approximation of \citet{zholobenko2024tokamak}. 
The effective parallel thermal conductivity is obtained
from the harmonic interpolation
\begin{equation}
\kappa_{\parallel,j}^{\mathrm{eff}} = \frac{\kappa_{\parallel,j}^{\mathrm{SH}}}{1+ \dfrac{\kappa_{\parallel,j}^{\mathrm{SH}}}{\alpha_j n_j v_{\mathrm{th},j}q_{95}R}},
\label{eq:connection_length_effective_conductivity}
\end{equation}
where the prescribed values of $q_{95} = 4$ and $R = 2$~m are used in the present simulations. Unlike the local-gradient model, this limiter does not depend explicitly on $\left|\nabla_{\parallel}T_j\right|$, but instead limits the conductivity using the characteristic length $q_{95}R$. The connection-length model may offer performance benefits by avoiding dependence of the flux limiter on local gradient quantities, while also preventing unphysically long mean free paths \cite{chankin2006solps}.


A related connection-length limiter is employed in the BOUT++ six-field model, although the local safety factor $q$ is used rather than a prescribed constant value. Parameter scans by \citet{xia2015nonlinear} showed that ELM heat losses are sensitive to the free-streaming coefficient $\alpha_j$, with lower values reducing the parallel heat flux and improving agreement with experimental energy losses.

\begin{figure}[!ht]
\centering
\includegraphics[width=0.49\linewidth]{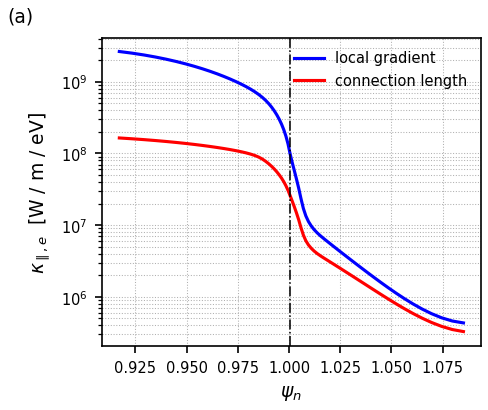}
\includegraphics[width=0.49\linewidth]{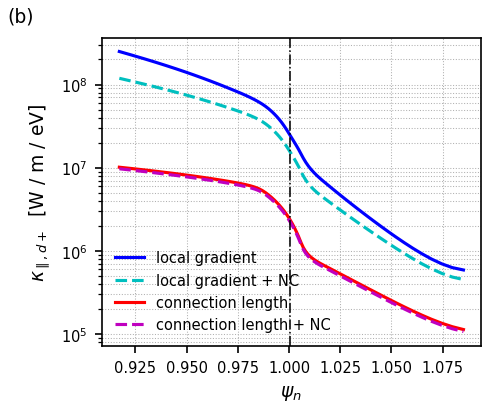}
\caption{Parallel thermal conductivities obtained with the local-gradient (Equation~\eqref{eq:local_gradient_effective_conductivity}) and connection-length (Equation~\eqref{eq:connection_length_effective_conductivity}) heat-flux limiters at the outer midplane: (a) electron thermal conductivity, $\kappa_{\parallel,e}$, and (b) deuterium-ion thermal conductivity, $\kappa_{\parallel,d+}$. The ion-conductivity panel also includes cases with the neoclassical correction, denoted by NC. All conductivity coefficients were evaluated under the same plasma conditions. Both panels use logarithmic vertical scales. }
\label{fig:kappa_parallel_limiter_comparison}
\end{figure}

Figure~\ref{fig:kappa_parallel_limiter_comparison} compares the outer-midplane electron and ion parallel thermal conductivities obtained with the local-gradient and connection-length limiters under identical plasma conditions. The electron conductivity is approximately an order of magnitude larger than the ion conductivity; for comparable temperatures, $\kappa_{\parallel,e}^{\mathrm{SH}}/\kappa_{\parallel,d+}^{\mathrm{SH}} \sim \sqrt{m_{d+}/m_e} \gg 1$. The connection-length limiter reduces the conductivities by approximately one order of magnitude in the pedestal region for both species. The difference between the two limiter models decreases in the colder, more collisional divertor region, where the effective conductivity approaches the Spitzer--Härm value.

\begin{figure}[!ht]
\centering
\includegraphics[width=0.49\linewidth]{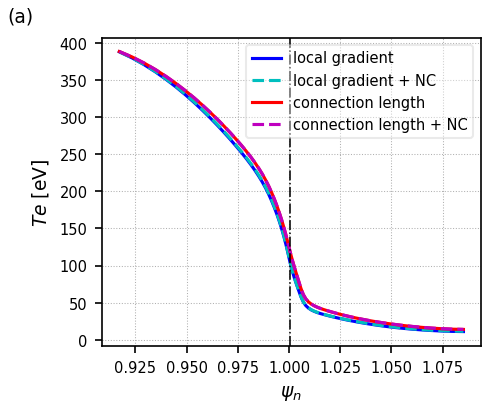}
\includegraphics[width=0.49\linewidth]{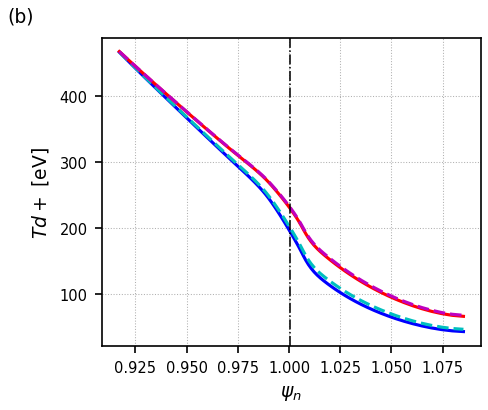}
\caption{Steady-state outer-midplane profiles of (a) electron temperature, $T_e$, and (b) deuterium-ion temperature, $T_{d+}$, comparing the local-gradient (Equation~\eqref{eq:local_gradient_effective_conductivity}) and connection-length (Equation~\eqref{eq:connection_length_effective_conductivity}) heat-flux limiter models. NC denotes the neoclassical correction.}
\label{fig:fluxLimiter_prof_OMP}
\end{figure}

\begin{figure}[!ht]
\centering
\includegraphics[width=0.49\linewidth]{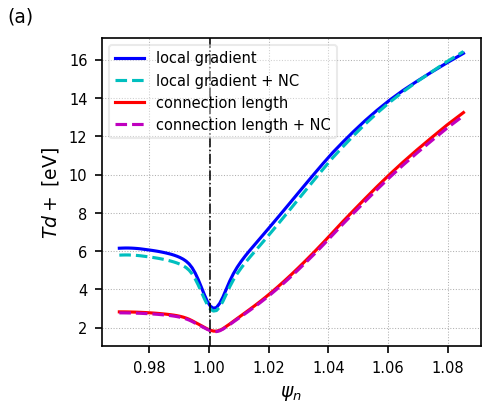}
\includegraphics[width=0.49\linewidth]{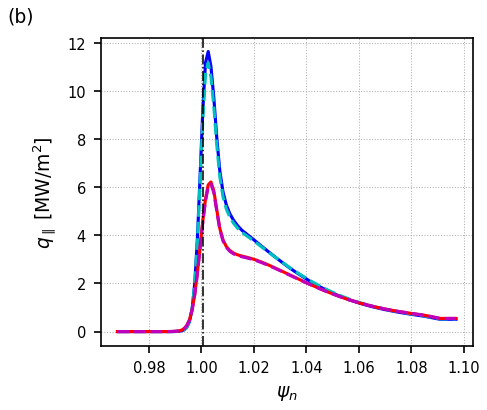}
\caption{Steady-state outer-target profiles of (a) deuterium-ion temperature, $T_{d+}$, and (b) total parallel heat flux, $q_{\parallel}$, comparing the local-gradient (Equation~\eqref{eq:local_gradient_effective_conductivity}) and connection-length (Equation~\eqref{eq:connection_length_effective_conductivity}) heat-flux limiter models. NC denotes the neoclassical correction.}
\label{fig:fluxLimiter_prof_OT}
\end{figure}

Figure~\ref{fig:fluxLimiter_prof_OMP} compares the outer-midplane electron and ion temperature profiles obtained with the two heat-flux limiter models. Despite the large differences in $\kappa_{\parallel,j}$, $T_e$ changes only slightly, while the connection-length limiter predicts $T_{d+}$ values approximately 25\% higher at the separatrix. This weak sensitivity can be partly attributed to adjustments in the parallel temperature gradient. Similarly, \citet{zholobenko2024tokamak} found no significant differences in the upstream plasma profiles obtained with free-streaming and Landau-fluid \citep{dimits2013fast} closures.

Figure~\ref{fig:fluxLimiter_prof_OT} compares the steady-state outer-target profiles obtained with the local-gradient and connection-length heat-flux limiter models. Panels (a) and (b) show $T_{d+}$ and the total parallel heat flux, $q_\parallel$, respectively. The connection-length limiter, which predicts a lower $\kappa_{\parallel,j}$, results in a lower target $T_{d+}$ and reduces $q_\parallel$ by approximately a factor of two. This behavior can be partly attributed to reduced parallel conductive transport, which increases the temperature drop along the divertor leg and allows volumetric energy losses to remove a larger fraction of the power before it reaches the target.  

Connection-length limiters have primarily been applied in 3D turbulence and ELM simulations. In addition to modifying the physical closure, limiting the parallel conductivity can reduce the fast conduction timescale and improve numerical performance. At high temperatures, parallel thermal conduction can strongly restrict the time step through the pressure equations \citep{zhu2018gdb}, particularly when CVODE is used. This restriction is generally dominated by electron conduction because of the much larger electron thermal conductivity, as discussed in relation to Figure~\ref{fig:kappa_parallel_limiter_comparison}. The connection-length formulation may therefore be particularly relevant for computationally demanding 3D simulations.

A neoclassical correction to the ion parallel thermal conductivity was also implemented in Hermes-3 following \citet{rozhansky2009new}:
\begin{equation}
\kappa_{\parallel,d+}^{\mathrm{NC}} = \kappa_{\parallel,d+}^{\mathrm{SH}} \frac{1.6\epsilon^{3/2}}{K_2}, 
\label{eq:neoclassical_ion_conductivity}
\end{equation}
where
\begin{equation} 
K_2 =  \frac{0.66 + 1.88\sqrt{\epsilon}-1.54\epsilon} {1+1.03\left(\sqrt{2}\nu_*\right)^{1/2}+0.31\sqrt{2}\nu_*} + \frac{1.17\epsilon^3\sqrt{2}\nu_*} {1+0.74\epsilon^{3/2}\sqrt{2}\nu_*}.
\label{eq:neoclassical_K2}
\end{equation}
and
\begin{equation}
\nu_* = \frac{L_c\nu_{d+}} {\epsilon^{3/2}\sqrt{2T_{d+}/m_{d+}}}.
\label{eq:neoclassical_collisionality}
\end{equation}
Here, $\epsilon$ is the inverse aspect ratio and $L_c$ is approximated as $L_c \simeq q_{95}R$, while the values of $q_{95} = 4$, $R = 2$~m, and $\epsilon = 0.3$ are treated as spatially constant model parameters. The correction is applied only to the ion conductivity and replaces $\kappa_{\parallel,d+}^{\mathrm{SH}}$ in the selected heat-flux limiter. It represents a correction to the ion parallel conductivity rather than a complete neoclassical transport model.

Figures~\ref{fig:kappa_parallel_limiter_comparison}, \ref{fig:fluxLimiter_prof_OMP}, and \ref{fig:fluxLimiter_prof_OT} include results obtained with and without the neoclassical ion-conductivity correction, denoted by NC. For the plasma conditions considered here, the correction produces negligible changes in the plasma profiles and outer-target heat flux. Its effect is expected to become more important at higher temperatures and in domains extending farther into the confined plasma.

\section{Conclusions} \label{sec:Conclusions}

This work presented a two-dimensional fluid model implemented in BOUT++/Hermes-3 for investigating detached divertor plasmas. The model incorporates impurity radiation, ionization, recombination, recycling, and neutral transport. Particular emphasis was placed on improving the PETSc SNES-based steady-state solution strategy for two-dimensional axisymmetric transport simulations with fluid neutrals.

Different preconditioner settings and alternative preconditioners were tested, and their relative performance was evaluated. The study showed that solver performance is strongly dependent on both preconditioner choice and parameter settings. Optimized \HYPRE{} ILU(k) and ILU(t) configurations substantially improved convergence compared with the default settings, while STRUMPACK used as an ILU preconditioner provided the most favorable performance for the cases considered here. STRUMPACK reached comparable simulation times approximately five times faster than the optimized \HYPRE{} configurations because of its reduced linear-to-nonlinear iteration ratio. These developments establish a foundation for subsequent multispecies simulations.

The optimized solver was then employed to characterize detached divertor conditions in DIII-D shot $\#174287$ at 4000~ms (large Type-I ELM regime). Experimental upstream density and temperature profiles were reproduced by adjusting the anomalous cross-field transport coefficients. Detachment was modeled using a nitrogen fixed-fraction radiation model with three impurity fractions, 1\%, 3\%, and 6\% N$_2$. While all cases reproduced similar upstream conditions, increasing the impurity fraction enhanced divertor radiation losses, reduced the target electron temperature below 5~eV, and lowered the peak heat flux, consistent with detached divertor conditions.

The steady-state particle and energy balances were examined by decomposing the density and pressure equations into their individual transport and source terms. This analysis identified the dominant mechanisms governing particle and energy transport in different plasma regions. Under the conditions considered here, ionization is the dominant particle source at the outer target and is balanced primarily by parallel advection, whereas recombination becomes important at the more strongly detached inner target. Electron energy losses in the outer divertor are dominated by ionization and impurity radiation, while ion energy losses occur primarily through collisional energy transfer to electrons.

To examine the overall detachment trends, an upstream separatrix density scan was performed. Increasing $n_{e,\mathrm{sep}}$ reduced the electron temperature at the divertor target. At higher densities, the integrated ion saturation current exhibited a rollover and subsequently decreased despite further increases in $n_{e,\mathrm{sep}}$, consistent with the progression toward detachment.

A connection-length parallel heat-flux limiter was implemented in Hermes-3 and compared with the local-gradient limiter to assess the sensitivity of the detached-divertor solution to the heat-flux-limiting model. The connection-length limiter reduced the effective parallel thermal conductivities, with only modest changes in the upstream temperature profiles, but lowered the outer-target parallel heat flux by about a factor of two. The neoclassical correction to the ion thermal conductivity had little effect under the conditions considered.

Several extensions are planned for future work. In particular, drift effects and impurity transport are already implemented, and future work will apply these capabilities to the present plasma model, together with the incorporation of molecular physics and additional ion and impurity species.The optimized steady-state solutions developed here also provide suitable background states for subsequent three-dimensional BOUT++ six-field simulations of ELM dynamics in detached plasmas.


\section*{Acknowledgement}
This work was performed under the auspices of the U.S. Department of Energy by Lawrence Livermore National Laboratory under Contract No. DE-AC52-07NA27344, LLNL-JRNL-2023188 and was supported by the SciDAC ABOUND Project (SCW1832). This work has been part funded by the Fusion Futures Programme. As announced by the UK Government in October 2023, Fusion Futures aims to provide holistic support for the development of the fusion sector. This material is based upon work supported by the U.S. Department of Energy, Office of Science, Office of Fusion Energy Sciences, using the DIII-D National Fusion Facility, a DOE Office of Science user facility, under Award LLNL-led SciDAC ABOUND Project SCW1832. This research used resources of the National Energy Research Scientific Computing Center (NERSC), a Department of Energy User Facility (project m4491-2025). We thank Dr. Zeyu Li for providing the kinetic-EFIT files used in this work and Dr. Renato Perillo for useful discussions. 




\bibliographystyle{abbrvnat}
\bibliography{Bibliography/References.bib}
\addcontentsline{toc}{section}{References}  

\end{document}